\documentclass[10pt]{iopart}
\pdfoutput=1 
\usepackage{graphicx}
\usepackage[nomarkers,nolists]{endfloat}

\begin{document}

\title[Theory of ELM Suppression by RMPs in DIII-D ]{Theory of Edge Localized Mode Suppression by Static Resonant Magnetic Perturbations in the DIII-D Tokamak}

\author{Richard Fitzpatrick}

\address{Institute for Fusion Studies,  Department of Physics,  University of Texas at Austin, 
Austin TX, 78712, USA}

\ead{rfitzp@farside.ph.utexas.edu}

\begin{abstract}
  The plasma response  to an externally generated, static, $n=2$, resonant
  magnetic perturbation (RMP)  in the pedestal region of DIII-D discharge \#158115 is investigated. In this particular discharge,
  the resonant amplitudes of the RMP are modulated
  in a cycloidal manner at a frequency of 1 Hz. Adopting the plausible hypothesis that mode penetration at the
  top of the pedestal is a necessary and sufficient condition for the RMP-induced suppression of edge localized modes (ELMs), recent
  cylindrical, nonlinear, reduced-magnetohydrodynamical (MHD) simulations performed by Hu, Nazikian {\it et al.} (2019) can account, in a
  quantitative fashion, for the density-pump out and RMP-induced ELM suppression threshold observed in DIII-D discharge \#158115.
  The primary aim of this paper is to employ analytic theory to further simplify the model of Hu, Nazikian {\it et al.} in such a
  manner that a complete simulation of  RMP-induced ELM-suppression in a DIII-D H-mode discharge  can be performed in a matter of minutes of real time.
  A secondary aim is to gain a more exact understanding of the physical mechanism that underlies RMP-induced ELM suppression in the
  DIII-D tokamak.

  We find that the response of the plasma to the applied RMP, in the immediate  vicinity of a given rational surface,
  is governed by {\em nonlinear}, rather than linear, theory. This is the case because the magnetic island widths associated with driven
  reconnection exceed the linear layer widths, even in cases where driven reconnection is strongly suppressed by plasma rotation.
  We find that the natural frequency at  a given rational surface (i.e., the helical frequency at which the locally resonant component of the RMP
  would need to propagate in order to maximize driven reconnection) is offset from the local ${\bf E}\times{\bf B}$ frame
  in the {\em ion}\/ diamagnetic direction. The size of the offset is mostly determined by neoclassical poloidal flow damping.
  Finally, our analytic nonlinear response model is found to be largely consistent with the simulations of Hu, Nazikian {\it et al}, and
  also correctly predicts the RMP-induced mode penetration threshold in DIII-D discharge
  \#158115. 
\end{abstract}
\submitted{\NF}
\maketitle

\section{Introduction}\label{s1}
Tokamak discharges operating in high-confinement mode (H-mode) \cite{wagner} exhibit intermittent bursts of heat and particle
transport, 
emanating from the outer regions of the plasma, that are known as {\em type-I edge localized modes} (ELMs) \cite{zohm}.
ELMs are fairly harmless in  present-day tokamaks possessing carbon plasma-facing components. However, large ELMs can cause
a problematic  influx of tungsten ions into the plasma core in   tokamaks possessing tungsten plasma-facing components \cite{den}. 
Moreover, it is estimated that the heat load that ELMs
will deliver to the tungsten plasma-facing components in a reactor-scale tokamak  will be large enough to cause
massive tungsten ion influx into the core, and that the erosion associated with this process  will 
unacceptably limit the lifetimes of these components  \cite{loarte}. Consequently, developing robust and effective
methods for ELM control is a high priority for the international magnetic fusion program. 

The most promising method for the control of ELMs is via the application of static  {\em resonant magnetic perturbations}\/  (RMPs).
Complete RMP-induced
ELM suppression was first demonstrated on the DIII-D tokamak \cite{evans}. Subsequently, either mitigation or compete suppression of
ELMs has been demonstrated on the JET \cite{jet}, ASDEX-U \cite{asdex}, KSTAR \cite{kstar}, 
and EAST \cite{east} tokamaks.

ELMs are  thought to
be caused by peeling-ballooning instabilities, with intermediate toroidal mode numbers, that are
driven by the strong pressure gradients and current density gradients characteristic of the edge region of an H-mode
tokamak discharge \cite{conner},
which is known as the {\em pedestal}\/ region.
The initial observations of RMP-induced ELM suppression were interpreted as an indication that the magnetic field in the pedestal 
is  rendered stochastic by the applied RMP, leading to greatly enhanced transport via thermal diffusion along
magnetic field-lines \cite{evans,fenstermacher}. This explanation was quickly abandoned because no significant 
reduction in the electron temperature gradient in the pedestal is observed during RMP-induced  ELM suppression experiments,
whereas a very significant reduction would be expected in the presence of stochastic fields. It is now generally accepted that
response currents generated within the pedestal, as a consequence of plasma rotation,  play a crucial role in the perturbed equilibrium in the
presence of RMPs, and that these currents  act to prevent the formation of RMP-driven magnetic island chains---a process known as
{\em shielding}---and,
thereby, significantly reduce the stochasticity of
the magnetic field \cite{berc}.

This paper concentrates on a particular (but completely typical) DIII-D H-mode discharge (\#158115 \cite{d158115}) in which ELMs were successfully
suppressed by an
externally applied $n=2$ RMP. In this discharge, the relative phase of magnetic perturbations generated by two sets of external
field coils is modulated sinusoidally at a frequency of 1 Hz, causing the amplitudes of the helical 
harmonics of the applied RMP that resonate in the pedestal region 
to modulate in a cycloidal manner at the same frequency. The application of the RMP is observed to generate two distinct
plasma responses. The first response---which is known as the {\em density pump-out}---is characterized by a reduction in the pedestal
density, accompanied by a much smaller reduction in the pedestal temperature, whose magnitude varies smoothly with the amplitude
of the edge-resonant components of the RMP. The density pump-out is not observed to be associated with ELM suppression. The second response---which is
known as {\em mode penetration}---occurs
when the amplitude of the edge-resonant harmonics of the RMP exceeds a certain critical value,  is associated with the sudden formation of
a locked magnetic island chain at a rational surface that lies close to the top of the pedestal, and is
accompanied by a sudden shift in the edge ion toroidal rotation. Mode penetration is observed to be strongly correlated  with ELM suppression.
If the amplitude of the edge-resonant harmonics of the RMP falls below
a second, somewhat smaller, critical value then  the locked magnetic island chain is expelled from the plasma, and the edge  ion toroidal
rotation  returns to its original value---this process is known as {\em mode unlocking}. Mode unlocking is invariably accompanied  by the
resumption of ELMs. Note that we can be sure that mode penetration is associated with the formation of a locked magnetic island
chain because when mode unlocking occurs the chain is observed to spin-up and decay away (see Figure~5 in \cite{d158115}). The fact that
mode penetration is a {\em necessary}\/ condition for ELM suppression is a reasonable inference from existing DIII-D experimental data \cite{d158115}. It is the
hypothesis of
this paper that mode penetration is also a {\em sufficient}\/ condition. Incidentally, it is not difficult to understand why the formation of
a locked island chain close to the top of the pedestal might give rise to ELM suppression. The flattening of the temperature and density
profiles across the island region reduces the pressure gradient at the top of the pedestal, which is likely to move the plasma
farther from the peeling-ballooning stability threshold. Moreover, the formation of a locked island is likely to enhance any
nonlinear interaction between the applied RMP and the peeling-ballooning mode (essentially by allowing the associated  magnetic fields
to phase lock to one another).

A complete numerical simulation of DIII-D discharge \#158115 would entail running a nonlinear extended (because the code would need to
incorporate diamagnetic and neoclassical effects) full-magnetohyrodynamical (MHD) (because ELMs cannot be modeled using reduced-MHD)  code in toroidal geometry (because ELMs cannot be
modeled in cylindrical geometry) for approximately $10^{\,8}$ Alfv\'{e}n times (which corresponds to 1 second of experimental time). Unfortunately,
this is completely impossible. (The most advanced, current-day,  nonlinear, toroidal, extended-MHD codes are typically only capable of running for
$10^{\,4}$ Alfv\'{e}n times
\cite{orain}.) Hence, to make further progress, some sort of reduced model is required. If we accept that mode penetration is a
sufficient condition for the achievement of ELM suppression then the problem is greatly simplified, because we do not need to
directly model ELMs, and, thus, we can perform a nonlinear reduced-MHD calculation in cylindrical geometry. 
This is precisely the approach taken in a recent paper by Hu, Nazikian {\em et al.} \cite{hu}. In this paper, computer simulations---made using the cylindrical, multi-harmonic, five-field, nonlinear,
initial-value code, TM1 \cite{tm1,tm2,tm3}---of   DIII-D discharge  \#158115 find
that the formation of RMP-driven magnetic
island chains at the bottom and the top of the pedestal can account, in a quantitative fashion, for the observed 
density pump-out, as well as the mode-penetration-induced ELM suppression threshold \cite{hu}. The simulations also
find that driven magnetic reconnection is strongly shielded in the middle of the pedestal. 
(Incidentally, the five fields in TM1 are
the poloidal magnetic
flux, the electron density, the perpendicular ion vorticity, the parallel ion velocity, and the electron temperature.)
It should be noted that the
simulations presented in \cite{hu} use the experimental density, temperature,  safety-factor, and radial electric field profiles, apply an appropriately modulated
$n=2$ RMP with the experimental spectrum of resonant harmonics, and simulate the response of the plasma for the
equivalent of 1 second of experimental time. 

Unfortunately, even the reduced model of \cite{hu} requires many tens of hours of cpu time (corresponding to many days of actual time) to perform a
complete simulation. 
It is the primary aim of this paper  to employ analytic theory to further simplify the model in such a manner that a complete simulation can
be performed in a matter of minutes in real time. 
A secondary aim is to gain a more exact understanding of the physical
mechanism that underlies RMP-induced ELM
suppression in the DIII-D tokamak. Two analytic theories 
are employed in this study. The first is the cylindrical, single-harmonic, four-field, {\em linear}, resonant
plasma response model of \cite{rfw,colef}. 
The second is the cylindrical, single-harmonic, four-field, {\em nonlinear}, resonant plasma response model of  \cite{rfx,rfy,rff}.
(Note that, unlike the
model used in \cite{hu}, for the sake of simplicity, the models used in this paper do not evolve the electron temperature profile
separately from the density profile. Hence, they are four-field, rather than five-field, models.)

\section{Preliminary Analysis}
\subsection{Plasma Equilibrium}
Consider a large aspect-ratio, low-$\beta$, tokamak plasma whose equilibrium magnetic flux surfaces map out (almost)
concentric circles in the poloidal plane. Such a plasma is well approximated as a periodic cylinder. Suppose that the
minor radius of the plasma is $a$. Standard right-handed cylindrical coordinates ($r$, $\theta$, $z$) are adopted. The system is
assumed to be periodic in the $z$-direction, with periodicity length $2\pi\,R_0$, where $R_0\gg a$ is the
simulated plasma  major radius. It is convenient to define the simulated toroidal angle $\phi=z/R_0$. 

The equilibrium magnetic field is written ${\bf B}({\bf r}) = [0,\, B_\theta(r),\, B_\phi]$. The associated equilibrium plasma
current density takes the form ${\bf j}({\bf r})= [0,\, 0,\, j_\phi(r)]$, where
\begin{equation}
\mu_0\,j_\phi(r )= \frac{1}{r}\,\frac{d(r\,B_\theta)}{dr}.
\end{equation}
The  safety factor,
\begin{equation}
q(r)=\frac{r\,B_\phi}{R_0\,B_\theta},
\end{equation}
parameterizes the helical pitches of equilibrium magnetic field-lines. In  a conventional tokamak plasma,
$|q(r)|$ is of order unity, and is a monotonically increasing function of $r$. 

\subsection{Plasma Response}
Consider the response of the plasma to a  static RMP.  Suppose that the RMP
has $|m|$ periods in the poloidal direction, and
$n>0$ periods in the toroidal direction. (Note that $m$ is positive if $q$ is positive, and vice versa.) It is convenient to express the perturbed magnetic field and the perturbed
plasma current density in terms of a  magnetic flux-function, $\psi(r,\theta,\phi,t)$. Thus,  
\begin{eqnarray}
\delta {\bf B} &=&\nabla\psi\times {\bf e}_z,\\[0.5ex]
\mu_0\,\delta{\bf j} &=& -\nabla^{\,2}\psi\,{\bf e}_z,
\end{eqnarray}
where 
\begin{equation}
\psi(r,\theta,\phi,t)= \hat{\psi}(r,t)\,\exp[\,{\rm i}\,(m\,\theta-n\,\phi)].
\end{equation}
This representation is valid provided that
$|m|/n\gg a/R_0$ \cite{rf1}.

As is well known, the response of the plasma to the applied  RMP is governed by the equations of
perturbed, marginally-stable (i.e., $\partial/\partial t \equiv 0$), {\em ideal magnetohydrodynamics}\/ (MHD) everywhere in the plasma, apart
from a relatively narrow  (in $r$) region   in the vicinity of the  rational surface, minor radius $r_s$, where
$q(r_s)=m/n$ \cite{rf1}.

It is convenient to parameterize the RMP in terms of the so-called {\em vacuum flux}, ${\mit\Psi}_v(t)= |{\mit\Psi}_v|\,{\rm e}^{-{\rm i}\,\varphi_v}$,
which is defined to be the value of $\hat{\psi}(r,t)$ at radius $r_s$ in the presence of the RMP, but in the absence of the plasma. Here,
$\varphi_v$ is the helical phase of the RMP, and is assumed to be constant in time.  Likewise, the response of the plasma in the vicinity of the
rational surface to the RMP is parameterized in terms of the so-called {\em reconnected flux}, ${\mit\Psi}_s(t)= |{\mit\Psi}_s|\,{\rm e}^{-{\rm i}\,\varphi_s}$,
which is the actual value of $\hat{\psi}(r,t)$ at radius $r_s$. Here, $\varphi_s(t)$ is the helical phase of the reconnected flux. 

The intrinsic stability of the $m$/$n$ tearing mode is governed by the {\em tearing stability index}\/  \cite{fkr}, 
\begin{equation}
{\mit\Delta}' = \left[\frac{d\ln\hat{\psi}}{dr}\right]_{r_s-}^{r_s+},
\end{equation}
where $\hat{\psi}(r)$ is a solution of the marginally-stable, ideal-MHD equations,  for the case of an $m$/$n$ helical perturbation, that satisfies physical boundary conditions at $r=0$ and $r=a$ (in the
absence of the RMP). According to resistive-MHD theory \cite{fkr,ruth}, if ${\mit\Delta}'>0$ then  the $m$/$n$ tearing mode spontaneously reconnects magnetic flux at the rational
surface to form a helical magnetic island chain. In the following, it is assumed that ${\mit\Delta}'<0$, so that the $m$/$n$ tearing mode is intrinsically stable. In this
situation, any magnetic reconnection that takes place at the rational surface is due solely to the RMP. 

\subsection{Linear Response Regime}\label{linear}
In this paper, we shall examine two different plasma response regimes at the rational surface. The first of these is the so-called  {\em semi-collisonal regime}\/ \cite{colef,drake,wael}.
This is a  linear, two-fluid, low-collisionality regime in which the reconnected magnetic flux induced  by the RMP is governed by \cite{rfx}
\begin{equation}\label{e9}
\frac{\delta_{\rm SC}}{r_s}\,\tau_R\left(\frac{d}{dt}+{\rm i}\,\omega\right){\mit\Psi}_s = {\mit\Delta}'\,r_s\,{\mit\Psi}_s + 2\,|m|\,{\cal A}\,{\mit\Psi}_v.
\end{equation}
Here,
\begin{equation}
\delta_{\rm SC} = \pi\,\frac{|n\,\omega_{\ast\,e}|^{1/2}\,\tau_H}{(\rho_s/r_s)\,\tau_R^{\,1/2}}\,r_s
\end{equation}
is the  linear layer width, whereas
\begin{eqnarray}
\tau_H& =& \frac{R_0}{|B_\phi|}\,\frac{\sqrt{\mu_0\,\rho(r_s)}}{n\,s},\\[0.5ex]
\tau_R &=&\mu_0\,r_s^{\,2}\,\sigma(r_s),\\[0.5ex]
\omega_{\ast\,e} &=& \frac{(dp_e/dr)_{r_s}}{e\,n_e(r_s)\,R_0\,B_\theta(r_s)},\\[0.5ex]
\rho_s &= &\frac{\sqrt{m_i\,T_e(r_s)}}{e\,|B_\phi|},
\end{eqnarray}
are the   hydromagnetic timescale,  resistive diffusion timescale, electron diamagnetic frequency, and ion sound radius, respectively, at the rational surface. Moreover,
$s=(d\ln q/d\ln r)_{r=r_s}$ is the local magnetic shear, ${\cal A}$ the amplification factor (i.e., the factor by which the radial
magnetic field at the rational surface due to the RMP is enhanced with respect to its vacuum value due to equilibrium plasma
currents external to the rational surface), 
$m_i$ the ion mass, and $e$ the magnitude of the electron
charge. Furthermore, $\sigma(r)$,   $n_e(r)$, $\rho(r)\equiv m_i \,n_e(r)$, $T_e(r)$,  $p_e(r)\equiv n_e(r)\,T_e(r)$ are the equilibrium plasma electrical conductivity, electron number density, mass density, 
electron temperature, and  electron pressure profiles, respectively.  Finally, 
\begin{equation}\label{e14}
\omega(t) = m\,{\mit\Omega}_{\theta}(r_s,t) - n\,{\mit\Omega}_{\phi}(r_s,t),
\end{equation}
where ${\mit\Omega}_{\theta}(r,t)$ and ${\mit\Omega}_{\phi}(r,t)$ are the   plasma poloidal and toroidal angular velocity profiles, respectively. [To be more exact,  ${\mit\Omega}_{\theta}(r,t)$ and ${\mit\Omega}_{\phi}(r,t)$ are
the poloidal and toroidal angular velocity profiles of an imaginary fluid that convects reconnected magnetic flux
at rational surfaces. It is assumed that changes in these velocity profiles are mirrored by changes in the actual
plasma velocity profiles. A magnetic island convected by the imaginary fluid propagates at its so-called natural frequency. The relationship between the
natural frequency and the ${\bf E}\times {\bf B}$ frequency is specified in Section~\ref{nat}.] It is helpful to define the viscous diffusion timescale at the rational surface,
\begin{equation}
\tau_V = \frac{r_s^{\,2}\,\rho(r_s)}{\mu(r_s)},
\end{equation}
where $\mu(r)$ is the equilibrium plasma (perpendicular) viscosity profile. 

It should be noted that the analysis of \cite{colef}, combined with the experimental
data listed in Table~\ref{table1}, leads to the conclusion that the appropriate cylindrical, four-field, linear, plasma response regime at the 
rational surface is the
so-called SCi (first semi-collisional)  regime. [In particular, the dimensionless parameters $c_\beta$, $D$, $P$, and $Q$ that
control the plasma response, according to the analysis of \cite{colef}, are calculated at the -8/2 and the -11/2
rational surfaces in the pedestal of DIII-D discharge \#158115 \cite{d158115} in Table~\ref{table1a}. In the first case, the fact that $D>1$   implies that we 
should consult Figure~3 in \cite{colef}. According to this figure, the fact that $c_\beta\,D<Q<D$ and $P>1$ 
 indicates the appropriate response regime at the -8/2 surface is the SCi regime. In the second case, the fact that $c_\beta^{\,1/3}< D < 1$
 implies that we should consult Figure~2  in \cite{colef}. According to this figure, the fact that $Q<D$ and $P>Q^{\,3}/D^{\,6}$ indicates the the appropriate response regime at the -11/2 surface is the SCi regime.
 The response and layer thickness in
the SCi regime are listed in Table~1 of \cite{colef}.]
Equation~(\ref{e9}) is a slightly
simplified implementation of the response of the plasma in the vicinity of the rational surface in the SCi  regime. 

Incidentally, the experimental data listed in Table~\ref{table1} is derived from the safety-factor, electron number density, electron temperature, and ${\bf E}\times {\bf B}$
frequency profiles shown in Figure~2 of \cite{hu}. The electron diamagnetic frequency, magnetic
shear, and $\eta_i$ values listed in the table are calculated directly from
these profiles. The  electron and ion temperature profiles are assumed to be the same. The values of the perpendicular momentum diffusivity are obtained  from the TRANSP code \cite{transp}.
The values of the effective ion charge number, $Z_{\rm eff}$, come from line emission spectroscopy. 

Equation~(\ref{e9}) can be  conveniently rewritten as an
{\em island width evolution equation}\/ \cite{rfx}, 
\begin{equation}
\frac{\delta_{\rm SC}}{r_s}\,\tau_R\,\frac{d}{dt}\!\left(\frac{W}{r_s}\right) = \frac{1}{2}\,\frac{W}{r_s}\left[{\mit\Delta}'\,r_s+2\,|m|\,{\cal A}\,\left(\frac{W_v}{W}\right)^{2}\cos\varphi\right],
\end{equation}
and an {\em island phase evolution equation},
\begin{equation}
\frac{\delta_{\rm SC}}{r_s}\,\tau_R\left(\frac{d\varphi}{dt}-\omega\right)= -2\,|m|\,{\cal A}\left(\frac{W_v}{W}\right)^2\sin\varphi.
\end{equation}
Here,
\begin{equation}
W(t) = 4\left(\frac{|{\mit\Psi}_s|}{s\,r_s\,|B_\theta(r_s)|}\right)^{1/2}r_s
\end{equation}
is the full (radial) width of the magnetic island chain that forms at the rational surface. (Incidentally, it is assumed that
$W\ll r_s$.)
Moreover,
\begin{equation}
W_v(t) = 4\left(\frac{|{\mit\Psi}_v|}{s\,r_s\,|B_\theta(r_s)|}\right)^{1/2}r_s
\end{equation}
is termed the vacuum island width. Finally,
\begin{equation}
\varphi(t) = \varphi_s(t)-\varphi_v
\end{equation}
is the helical phase of the island chain relative to the RMP. 

The semi-collisional  response regime holds when $W < \delta_{\rm SC}$. That is, when the magnetic
island width falls below the linear layer width. 

\subsection{Nonlinear Response Regime}\label{nonlinear}
The second response regime investigated in this paper is the so-called {\em Rutherford regime}\/ \cite{rf2}.
This is a nonlinear regime in which the reconnected magnetic flux induced  by the RMP is governed by two
equations. The first of these is the {\em Rutherford island width evolution equation}\/ \cite{ruth},
\begin{equation}\label{e17}
{\cal I}\,\tau_R\,\frac{d}{dt}\!\left(\frac{W}{r_s}\right) = {\mit\Delta}' \,r_s+ 2\,|m|\,{\cal A}\left(\frac{W_v}{W}\right)^2\cos\varphi,
\end{equation}
where ${\cal I}=0.8227$. 
 The second governing equation is the so-called {\em no-slip constraint}\/ \cite{rf1}, 
\begin{equation}\label{e20}
\frac{d\varphi_s}{dt} - \omega=0,
\end{equation}
according to which the island chain is forced to co-rotate with the aforementioned (see Section~\ref{linear}) magnetic-flux-convecting imaginary fluid at the rational surface. 

The Rutherford response regime holds when 
$W> \delta_{\rm SC}$.
That is, when  the magnetic island width  exceeds the linear layer width.  

\subsection{Plasma Angular Velocity Evolution}\label{angular}
It is easily demonstrated that zero net electromagnetic torque can be exerted on magnetic flux surfaces located in a region of the
plasma that is governed by the equations of marginally-stable, ideal-MHD \cite{rf1}. Thus, any electromagnetic torque exerted on the plasma by the RMP
develops in the immediate vicinity of the rational surface, where ideal-MHD breaks down. The
net  poloidal and toroidal  electromagnetic torques  exerted in the vicinity of the rational surface by the RMP  take the forms \cite{rf1,rf2}
\begin{eqnarray}
T_{\theta\,{\rm EM}} &= &-\frac{4\pi^{\,2}\,|m|\,m\,R_0}{\mu_0}\,{\cal A}\,|{\mit\Psi}_v|\,|{\mit\Psi}_s|\,\sin\varphi,\\[0.5ex]
T_{\phi\,{\rm EM}} &= &\frac{4\pi^{\,2}\,|m|\,n\,R_0}{\mu_0}\,{\cal A}\,|{\mit\Psi}_v|\,|{\mit\Psi}_s|\,\sin\varphi,
\end{eqnarray}
respectively. 

We can write
\begin{eqnarray}\label{e22}
{\mit\Omega}_\theta(r,t) &=&{\mit\Omega}_{\theta\,0}(r) + {\mit\Delta\Omega}_\theta(r,t),\\[0.5ex]
{\mit\Omega}_\phi(r,t) &=&{\mit\Omega}_{\phi\,0}(r) + {\mit\Delta\Omega}_\phi(r,t),\label{e23}
\end{eqnarray}
where ${\mit\Omega}_{\theta\,0}(r)$ and ${\mit\Omega}_{\phi\,0}(r)$ are the equilibrium poloidal and toroidal
plasma angular velocity profiles, respectively, whereas ${\mit\Delta\Omega}_{\theta}(r,t)$ and ${\mit\Delta\Omega}_{\phi}(r,t)$ 
are the respective modifications to these profiles induced by the aforementioned electromagnetic torques. 
The modifications to the angular velocity profiles are governed by the poloidal and toroidal angular equations of
motion of the plasma, which take the respective forms \cite{rf1,hirsh}
\begin{eqnarray}\label{eom1}\fl
4\pi^{\,2}\,R_0\left[(1+2\,q^{\,2})\,\rho\,r^{\,3}\,\frac{\partial{\mit\Delta\Omega}_\theta}{\partial t}-\frac{\partial}{\partial r}\!\left(
\mu\,r^{\,3}\,\frac{\partial{\mit\Delta\Omega}_\theta}{\partial r}\right) +\rho\,r^{\,3}\,\frac{{\mit\Delta\Omega}_\theta}{\tau_\theta}\right] 
\nonumber\\[0.5ex]\phantom{====}
=T_{\theta\,{\rm EM}}\,\delta(r-r_s),\\[0.5ex]
\fl 4\pi^{\,2}\,R_0^{\,3}\left[\rho\,r\,\frac{\partial{\mit\Delta\Omega}_\phi}{\partial t}-\frac{\partial}{\partial r}\!\left(
\mu\,r\,\frac{\partial{\mit\Delta\Omega}_\phi}{\partial r}\right) +\rho\,r\,\frac{{\mit\Delta\Omega}_\phi}{\tau_\phi}\right]
=T_{\phi\,{\rm EM}}\,\delta(r-r_s),\label{eom2}
\end{eqnarray}
and are subject to the spatial boundary conditions \cite{rf1}
\begin{eqnarray}
\frac{\partial{\mit\Delta\Omega}_\theta(0,t)}{\partial r} &=&
\frac{\partial{\mit\Delta\Omega}_\phi(0,t)}{\partial r} =0,\\[0.5ex]
{\mit\Delta\Omega}_\theta(a,t)&=&{\mit\Delta\Omega}_\phi(a,t)=0.
\end{eqnarray}
Here, $\tau_\theta(r)$ is  the neoclassical poloidal flow-damping time profile \cite{stix}, 
and $\tau_\phi(r)$  the neoclassical toroidal flow-damping time profile. The neoclassical
toroidal flow-damping is assumed to be generated by non-resonant components of the applied
RMP \cite{shaing,cole}.  The factor $(1+2\,q^{\,2})$ in (\ref{eom1}) derives from the fact that incompressible poloidal
flow has a poloidaly-varying toroidal component that effectively increases the plasma mass being accelerated by the poloidal
flow-damping force \cite{hirsh}.
It turns out that, in the presence of strong poloidal and toroidal flow-damping, the
modifications to the plasma poloidal and toroidal angular velocity profiles are localized in the vicinity of the rational
surface \cite{cole}. Assuming that this is the case,  it is a good approximation to replace $q$, $\rho$, $\mu$, $\tau_\theta$, and $\tau_\phi$ in
 (\ref{eom1}) and (\ref{eom2}) by their values  at the rational surface. We are, nevertheless, assuming that the localization width greatly exceeds the
linear layer width (or the island width, in the nonlinear case).

Equations~(\ref{e14}), (\ref{e22}), and (\ref{e23}) imply that
\begin{equation}
\omega(t)=\omega_0 +m\,{\mit\Delta\Omega}_{\theta}(r_s,t)-n\,{\mit\Delta\Omega}_{\phi}(r_s,t),
\end{equation}
where 
\begin{equation}
\omega_0 = m\,{\mit\Omega}_{\theta\,0}(r_s)-n\,{\mit\Omega}_{\phi\,0}(r_s)
\end{equation}
is the so-called {\em natural frequency}\/ of the $m$/$n$ tearing mode. In other words,
$\omega_0$ is the helical phase velocity of a naturally unstable $m$/$n$ tearing mode in the
absence of the RMP.

\subsection{Natural Frequency}\label{nat}
According to the cylindrical, single-helicity, four-field, {\em linear}\/ analysis of \cite{colef}, the appropriate natural frequency for the
linear response model is
\begin{equation}
\omega_0 = - n\,(\omega_E+\omega_{\ast\,e}),
\end{equation}
where $\omega_E= E_r(r_s)/[R_0\,B_\theta(r_s)]$ is the ${\bf E}\times {\bf B}$ frequency at the rational surface, and $E_r(r)$
is the equilibrium radial electric field profile. 

According to the cylindrical, single-helicity, four-field, {\em nonlinear}\/  analysis of  \cite{rff}, the appropriate natural frequency for the
nonlinear response model is
\begin{equation}\label{enl1}
\omega_0 = -n\,\omega_E - n\left(1-\frac{\eta_i\,\lambda_{\theta\,i}}{1+\eta_i}\right)\omega_{\ast\,i},
\end{equation}
where
\begin{equation}\label{enl2}
\omega_{\ast\,i} = -\frac{(dp_i/dr)_{r_s}}{e\,n_e(r_s)\,R_0\,B_\theta(r_s)}
\end{equation}
is the ion diamagnetic frequency at the rational surface, $\eta_i=(d\ln T_i/d\ln n_e)_{r=r_s}$,  $T_i(r)$  the equilibrium ion temperature profile, $p_i(r)=n_e(r)\,T_i(r)$ the equilibrium ion pressure profile,  and the dimensionless parameter
$\lambda_{\theta\,i}$ is specified in \ref{appz1}. 

\section{Linear Response Model}\label{slinearx}
\subsection{Unnormalized Linear Response Model}
According to Sections~\ref{linear} and \ref{angular}, the complete linear response model takes the
form
\begin{eqnarray}
\frac{\delta_{\rm SC}}{r_s}\,\tau_R\,\frac{d}{dt}\!\left(\frac{W}{4\,r_s}\right)=\frac{1}{2}\left(\frac{W}{4\,r_s}\right)\left[{\mit\Delta}'\,r_s
+2\,|m|\,{\cal A}\left(\frac{W_v}{W}\right)^2\cos\varphi\right],\\[0.5ex]
\frac{\delta_{\rm SC}}{r_s}\,\tau_R\left(\frac{d\varphi}{dt}-\omega\right)=-2\,|m|\,{\cal A}\left(\frac{W_v}{W}\right)^2 \sin\varphi,\\[0.5ex]
 \omega = \omega_0+m\,{\mit\Delta\Omega}_\theta(r_s,t)-n\,{\mit\Delta\Omega}_\phi(r_s,t),\\[0.5ex]
\fl \left[(1+2\,q_s^{\,2})\,\rho\,r^{\,3}\,\frac{\partial{\mit\Delta\Omega}_\theta}{\partial t}-\frac{\partial}{\partial r}\!\left(
\mu\,r^{\,3}\,\frac{\partial{\mit\Delta\Omega}_\theta}{\partial r}\right) +\rho\,r^{\,3}\,\frac{{\mit\Delta\Omega}_\theta}{\tau_\theta}\right]
 \nonumber\\[0.5ex]
\phantom{==}=-\frac{|m|\,m}{\mu_0}\,{\cal A}\left(\frac{W_v}{4\,r_s}\right)^2\left(\frac{W}{4\,r_s}\right)^2\left[s\,r_s\,B_{\theta}(r_s)\right]^{\,2}\,\sin\varphi\,\delta(r-r_s),\\[0.5ex]
\fl R_0^{\,2}\left[\rho\,r\,\frac{\partial{\mit\Delta\Omega}_\phi}{\partial t}-\frac{\partial}{\partial r}\!\left(
\mu\,r\,\frac{\partial{\mit\Delta\Omega}_\phi}{\partial r}\right) +\rho\,r\,\frac{{\mit\Delta\Omega}_\phi}{\tau_\phi}\right]\nonumber\\[0.5ex]
\phantom{==}=\frac{|m|\,n}{\mu_0}\,{\cal A}\left(\frac{W_v}{4\,r_s}\right)^2\left(\frac{W}{4\,r_s}\right)^2 \left[s\,r_s\,B_{\theta}(r_s)\right]^{\,2}\,\sin\varphi\,\delta(r-r_s),\\[0.5ex]
\frac{\partial{\mit\Delta\Omega}_\theta(0,t)}{\partial r} =
\frac{\partial{\mit\Delta\Omega}_\phi(0,t)}{\partial r} =0,\\[0.5ex]
{\mit\Delta\Omega}_\theta(a,t)={\mit\Delta\Omega}_\phi(a,t)=0,
\end{eqnarray}
where $q_s=m/n$. 

\subsection{Normalized Linear Response Model}\label{norm}
It is helpful to define the typical {\em semi-collisional magnetic reconnection timescale},
\begin{equation}
\tau_{\rm SC} = \frac{\delta_{\rm SC}}{r_s}\,\frac{\tau_R}{2\,|m|}.
\end{equation}
Let
$\hat{r} = r/a$, 
$\hat{t} = t/\tau_{\rm SC}$,
$\hat{\omega}_0= \omega_0\,\tau_{\rm SC}$,
$\hat{\omega}_\theta(\hat{r},\hat{t}) = -m\,{\mit\Delta\Omega}_\theta(r,t)\,\tau_{\rm SC}$, 
$\hat{\omega}_\phi(\hat{r},\hat{t}) = n\,{\mit\Delta\Omega}_\phi(r,t)\,\tau_{\rm SC}$, 
$\hat{W}= W/\delta_{\rm SC}$,
$\hat{W}_v= W_v/\delta_{\rm SC}$.
The normalized linear response model reduces to 
\begin{eqnarray}
\frac{d\hat{W}}{d\hat{t}}= \frac{\hat{W}}{2}\left[-\hat{\mit\Delta}'+{\cal A}\,\left(\frac{\hat{W}_v}{\hat{W}}\right)^2\cos\varphi\right],\\[0.5ex]
\frac{d\varphi}{d\hat{t}}= \hat{\omega}-{\cal A}\left(\frac{\hat{W}_v}{\hat{W}}\right)^2\sin\varphi,\\[0.5ex]
\hat{\omega}= \hat{\omega}_0-\hat{\omega}_\theta(\hat{r}_s,\hat{t})-\hat{\omega}_\phi(\hat{r}_s,\hat{t}),\\[0.5ex]
\fl (1+2\,q_s^{\,2})\,\hat{r}^{\,3}\,\frac{\partial\hat{\omega}_\theta}{\partial\hat{t}}-\nu_\mu\,\frac{\partial}{\partial\hat{r}}\!\left(\hat{r}^{\,3}\,\frac{\partial\hat{\omega}_\theta}{\partial\hat{r}}\right)+\nu_\theta\,\hat{r}^{\,3}\,\hat{\omega}_\theta
 =\frac{ {\cal A}\,\hat{W}_v^{\,2}\,\hat{W}^{\,2}}{\hat{W}_0^{\,4}}\,\sin\varphi\,\delta(\hat{r}-\hat{r}_s),\label{e47}\\[0.5ex]
\fl \hat{r}\,\frac{\partial\hat{\omega}_\phi}{\partial\hat{t}}-\nu_\mu\,\frac{\partial}{\partial\hat{r}}\!\left(\hat{r}\,\frac{\partial\hat{\omega}_\phi}{\partial\hat{r}}\right)+\nu_\phi\,\hat{r}\,\hat{\omega}_\phi 
= \left(\frac{\epsilon_a}{q_s}\right)^2 \frac{{\cal A}\,\hat{W}_v^{\,2}\,\hat{W}^{\,2}}{\hat{W}_0^{\,4}}\,\sin\varphi\,\delta(\hat{r}-\hat{r}_s),\\[0.5ex]
\frac{\partial\hat{\omega}_\theta(0,\hat{t})}{\partial\hat{r}}= \frac{\partial\hat{\omega}_\phi(0,\hat{t})}{\partial\hat{r}}=0,\\[0.5ex]
\hat{\omega}_\theta(1,\hat{t})=\hat{\omega}_\phi(1,\hat{t})=0,\label{e50}
\end{eqnarray}
where $\hat{r}_s=r_s/a$, $\hat{\mit\Delta}'= (-{\mit\Delta}'\,r_s)/(2\,|m|)$, $\epsilon_a=a/R_0$, 
$\nu_\theta=\tau_{\rm SC}/\tau_\theta$,
$\nu_\phi = \tau_{\rm SC}/\tau_\phi$,
$\nu_\mu = (\tau_{\rm SC}/\tau_V)\,(r_s/a)^2$,
$\hat{W}_0 = W_0/\delta_{\rm SC}$, and 
\begin{equation}
W_0 = 4\left(2\,\frac{\tau_H^{\,2}}{\tau_{\rm SC}\,\tau_R}\,\frac{r_s}{\delta_{\rm SC}}\right)^{1/4}a.
\end{equation}

\subsection{Solution of Plasma Angular Equations of Motion}\label{angle}
We can solve (\ref{e47})--(\ref{e50}) by writing \cite{chapman}
\begin{eqnarray}
\hat{\omega}_\theta(\hat{r},\hat{t}) &=&\sum_{n=1,N} a_n(\hat{t})\,\frac{y_n(\hat{r})}{y_n(\hat{r}_s)},\\[0.5ex]
\hat{\omega}_\phi(\hat{r},\hat{t}) &=&\sum_{n=1,N} b_n(\hat{t})\,\frac{z_n(\hat{r})}{z_n(\hat{r}_s)},
\end{eqnarray}
where
\begin{eqnarray}
y_n(\hat{r}) &= &\frac{J_1(j_{1,n}\,\hat{r})}{\hat{r}},\\[0.5ex]
z_n(\hat{r}) &=& J_0(j_{0,n}\,\hat{r}).
\end{eqnarray}
The solution is exact in the limit $N\rightarrow\infty$. 
Here, $J_m(z)$ is a standard Bessel function, and $j_{m,n}$ denotes the $n$th zero of the $J_m(z)$ Bessel function \cite{abram1}.
It is easily demonstrated that
\begin{eqnarray}
\frac{d}{d\hat{r}}\!\left(\hat{r}^{\,3}\,\frac{dy_n}{d\hat{r}}\right)& =& - j_{1,n}^{\,2}\,\hat{r}^{\,3}\,y_n,\\[0.5ex]
\frac{d}{d\hat{r}}\!\left(\hat{r}\,\frac{dz_n}{d\hat{r}}\right) &= &- j_{0,n}^{\,2}\,\hat{r}\,z_n,
\end{eqnarray}
and \cite{grad2}
\begin{eqnarray}
\int_0^1\hat{r}^{\,3}\,y_n(\hat{r})\,y_m(\hat{r})\,d\hat{r}& =& \frac{1}{2}\left[J_2(j_{1,n})\right]^{\,2}\,\delta_{nm},\\[0.5ex]
\int_0^1\hat{r}\,z_n(\hat{r})\,z_m(\hat{r})\,d\hat{r}& = &\frac{1}{2}\left[J_1(j_{0,n})\right]^{\,2}\,\delta_{nm}.
\end{eqnarray}
Hence, we  obtain
\begin{eqnarray}
\hat{\omega}_\theta(\hat{r}_s,\hat{t})&=&\sum_{n=1,N} a_n(\hat{t}),\\[0.5ex]
\hat{\omega}_\phi(\hat{r}_s,\hat{t})&=& \sum_{n=1,N} b_n(\hat{t}),
\end{eqnarray}
where
\begin{eqnarray}
(1+2\,q_s^{\,2})\,\frac{d a_n}{d\hat{t}} + (\nu_\theta+\nu_\mu\,j_{1,n}^{\,2})\,a_n = \alpha_n(\hat{r}_s)\,\frac{{\cal A}\,\hat{W}_v^{\,2}\,\hat{W}^{\,2}}{\hat{W}_0^{\,4}}\,\sin\varphi,\\[0.5ex]
\frac{d b_n}{d\hat{t}} + (\nu_\phi+\nu_\mu\,j_{0,n}^{\,2})\,b_n =\epsilon\,\beta_n(\hat{r}_s)\,\frac{{\cal A}\,\hat{W}_v^{\,2}\,\hat{W}^{\,2}}{\hat{W}_0^{\,4}}\,\sin\varphi,
\end{eqnarray}			
and
\begin{eqnarray}
\alpha_n &= &\left[\frac{\sqrt{2}\,J_1(j_{1,n}\,\hat{r}_s)}{\hat{r}_s\,J_2(j_{1,n})}\right]^{2},\\[0.5ex]
\beta_n &=  &\left[\frac{\sqrt{2}\,J_0(j_{0,n}\,\hat{r}_s)}{J_1(j_{0,n})}\right]^{2},\\[0.5ex]
\epsilon &= &\left(\frac{\epsilon_a}{q_s}\right)^{2}.
\end{eqnarray}

\subsection{Final Form of  Normalized Linear Response Model}
The normalized linear model reduces to the following closed set of equations:
\begin{eqnarray}\label{ex1}
\frac{d\hat{W}}{d\hat{t}}= \frac{\hat{W}}{2}\left(-\hat{\mit\Delta}'+\frac{b_f}{\hat{W}^{\,2}}\,\cos\varphi\right),\\[0.5ex]
\frac{d\varphi}{d\hat{t}}= \hat{\omega}-\frac{b_f}{\hat{W}^{\,2}}\,\sin\varphi,\\[0.5ex]
(1+2\,q_s^{\,2})\,\frac{d a_n}{d\hat{t}} + (\nu_\theta+\nu_\mu\,j_{1,n}^{\,2})\,a_n = \alpha_n(\hat{r}_s)\,L\,b_f\,\hat{W}^{\,2}\,\sin\varphi,\\[0.5ex]
\frac{d b_n}{d\hat{t}} + (\nu_\phi+\nu_\mu\,j_{0,n}^{\,2})\,b_n = \epsilon\,\beta_n(\hat{r}_s)\,L\, b_f\,\hat{W}^{\,2}\,\sin\varphi,\\[0.5ex]
\hat{\omega} = \hat{\omega}_0-\sum_{n=1,N} a_n-\sum_{n=1,N} b_n,\label{ex2}
\end{eqnarray}
where
$b_f = {\cal A}\,\hat{W}_v^{\,2}$, and 
$L = \hat{W}_0^{-4}$.

If we define
\begin{eqnarray}
X&=&\hat{W}^{\,2}\,\cos\varphi,\label{exdef}\\[0.5ex]
Y&=&\hat{W}^{\,2}\,\sin\varphi\label{eydef}
\end{eqnarray}
then the previous set of equations yield 
\begin{eqnarray}\label{ast1}
\frac{d X}{d\hat{t}}= -\hat{\omega}\,Y - \hat{\mit\Delta}'\,X+b_{f},\\[0.5ex]
\frac{d Y}{d\hat{t}}=\hat{\omega}\,X-\hat{\mit\Delta}'\,\,Y,\\[0.5ex]
(1+2\,q_s^{\,2})\,\frac{d a_n}{d\hat{t}} + (\nu_\theta+\nu_\mu\,j_{1,n}^{\,2})\,a_n = \alpha_n(\hat{r}_s)\,L\,b_f\,Y,\\[0.5ex]
\frac{d b_n}{d\hat{t}} + (\nu_\phi+\nu_\mu\,j_{0,n}^{\,2})\,b_n = \epsilon\,\beta_n(\hat{r}_s)\,L\,b_f\,Y,\\[0.5ex]
\hat{\omega} = \hat{\omega}_0-\sum_{n=1,N} a_n-\sum_{n=1,N} b_n.\label{ast2}
\end{eqnarray}

\section{Nonlinear Response Model}\label{snonlinearx}
\subsection{Unnormalized Nonlinear Response Model}
According to Sections~\ref{nonlinear} and \ref{angular}, the complete nonlinear response model takes the form 
\begin{eqnarray}
{\cal I}\,\tau_R\,\frac{d}{dt}\!\left(\frac{W}{r_s}\right) = {\mit\Delta}' \,r_s+ 2\,|m|\,{\cal A}\left(\frac{W_v}{W}\right)^2\cos\varphi,\\[0.5ex]
\frac{d\varphi}{dt}=  \omega,\\[0.5ex]
 \omega = \omega_0+m\,{\mit\Delta\Omega}_\theta(r_s,t)-n\,{\mit\Delta\Omega}_\phi(r_s,t),\\[0.5ex]
 \fl \left[(1+2\,q_s^{\,2})\,\rho\,r^{\,3}\,\frac{\partial{\mit\Delta\Omega}_\theta}{\partial t}-\frac{\partial}{\partial r}\!\left(
\mu\,r^{\,3}\,\frac{\partial{\mit\Delta\Omega}_\theta}{\partial r}\right) +\rho\,r^{\,3}\,\frac{{\mit\Delta\Omega}_\theta}{\tau_\theta}\right]\nonumber\\[0.5ex]
\phantom{==}
=-\frac{|m|\,m}{\mu_0}\,{\cal A}\left(\frac{W_v}{4\,r_s}\right)^2\left(\frac{W}{4\,r_s}\right)^2 \left[s\,r_s\,B_{\theta}(r_s)\right]^{\,2}\,\sin\varphi\,\delta(r-r_s),\\[0.5ex]
\fl R_0^{\,2}\left[\rho\,r\,\frac{\partial{\mit\Delta\Omega}_\phi}{\partial t}-\frac{\partial}{\partial r}\!\left(
\mu\,r\,\frac{\partial{\mit\Delta\Omega}_\phi}{\partial r}\right) +\rho\,r\,\frac{{\mit\Delta\Omega}_\phi}{\tau_\phi}\right]\nonumber\\[0.5ex]
\phantom{==}=\frac{|m|\,n}{\mu_0}\,{\cal A}\left(\frac{W_v}{4\,r_s}\right)^2\left(\frac{W}{4\,r_s}\right)^2 \left[s\,r_s\,B_{\theta}(r_s)\right]^{\,2}\,\sin\varphi\,\delta(r-r_s),\\[0.5ex]
\frac{\partial{\mit\Delta\Omega}_\theta(0,t)}{\partial r} =
\frac{\partial{\mit\Delta\Omega}_\phi(0,t)}{\partial r} =0,\\[0.5ex]
{\mit\Delta\Omega}_\theta(a,t)={\mit\Delta\Omega}_\phi(a,t)=0.
\end{eqnarray}

\subsection{Normalized Nonlinear Response Model}
The normalized form of the nonlinear response model is
\begin{eqnarray}
{\cal I}\,\frac{d\hat{W}}{d\hat{t}}=-\hat{\mit\Delta}'+{\cal A}\left(\frac{\hat{W}_v}{\hat{W}}\right)^2\cos\varphi,\\[0.5ex]
\frac{d\varphi}{d\hat{t}}= \hat{\omega},\\[0.5ex]
\hat{\omega}= \hat{\omega}_0-\hat{\omega}_\theta(\hat{r}_s,\hat{t})-\hat{\omega}_\phi(\hat{r}_s,\hat{t}),\\[0.5ex]
\fl (1+2\,q_s^{\,2})\,\hat{r}^{\,3}\,\frac{\partial\hat{\omega}_\theta}{\partial\hat{t}}-\nu_\mu\,\frac{\partial}{\partial\hat{r}}\!\left(\hat{r}^{\,3}\,\frac{\partial\hat{\omega}_\theta}{\partial\hat{r}}\right)+\nu_\theta\,\hat{r}^{\,3}\,\hat{\omega}_\theta 
=\frac{{\cal A}\,\hat{W}_v^{\,2}\,\hat{W}^{\,2}}{\hat{W}_0^{\,4}}\,\sin\varphi\,\delta(\hat{r}-\hat{r}_s),\\[0.5ex]
\fl \hat{r}\,\frac{\partial\hat{\omega}_\phi}{\partial\hat{t}}-\nu_\mu\,\frac{\partial}{\partial\hat{r}}\!\left(\hat{r}\,\frac{\partial\hat{\omega}_\phi}{\partial\hat{r}}\right)+\nu_\phi\,\hat{r}\,\hat{\omega}_\phi 
= \left(\frac{\epsilon_a}{q_s}\right)^2\frac{{\cal A}\, \hat{W}_v^{\,2}\,\hat{W}^{\,2}}{\hat{W}_0^{\,4}}\,\sin\varphi\,\delta(\hat{r}-\hat{r}_s),\\[0.5ex]
\frac{\partial\hat{\omega}_\theta(0,\hat{t})}{\partial\hat{r}} = \frac{\partial\hat{\omega}_\phi(0,\hat{t})}{\partial\hat{r}}=0,\\[0.5ex]
\hat{\omega}_\theta(1,\hat{t})=\hat{\omega}_\phi(1,\hat{t})=0.
\end{eqnarray}

\subsection{Final Form of Normalized  Nonlinear Response Model}
The normalized nonlinear model reduces to the following closed set of equations:
\begin{eqnarray}\label{e94}
{\cal I}\,\frac{d\hat{W}}{d\hat{t}}=-\hat{\mit\Delta}'+\frac{b_f}{\hat{W}^{\,2}}\,\cos\varphi,\\[0.5ex]
\frac{d\varphi}{d\hat{t}}= \hat{\omega},\\[0.5ex]
(1+2\,q_s^{\,2})\,\frac{d a_n}{d\hat{t}} + (\nu_\theta+\nu_\mu\,j_{1,n}^{\,2})\,a_n =\alpha_n(\hat{r}_s)\,L\,b_f\,\hat{W}^{\,2}\,\sin\varphi,\label{e105}\\[0.5ex]
\frac{d b_n}{d\hat{t}} + (\nu_\phi+\nu_\mu\,j_{0,n}^{\,2})\,b_n =\epsilon\,\beta_n(\hat{r}_s)\,L\,b_f\,\hat{W}^{\,2}\,\sin\varphi,\label{e106}\\[0.5ex]
\hat{\omega} = \hat{\omega}_0-\sum_{n=1,N} a_n-\sum_{n=1,N} b_n.\label{e100}
\end{eqnarray}
Here, we have reused the analysis of Section~\ref{angle} to solve the plasma angular equations of motion.

If we express the previous equations in terms of the variables $X$ and $Y$, which are defined in (\ref{exdef}) and (\ref{eydef}), then we get 
\begin{eqnarray}
\frac{d X}{d\hat{t}}= -\hat{\omega}\,Y - f(X, Y)\,\hat{\mit\Delta}'\,X+
 f(X, Y)\left(\frac{X^{\,2}}{X^{\,2}+Y^{\,2}}\right)b_{f},\label{e103}\\[0.5ex]
\frac{d Y}{d\hat{t}}=\hat{\omega}\,X- f(X, Y)\,\hat{\mit\Delta}'\,\,Y+f(X, Y)\left(\frac{X\,Y}{X^{\,2}+Y^{\,2}}\right)b_{f},\label{e104}\\[0.5ex]
(1+2\,q_s^{\,2})\,\frac{d a_n}{d\hat{t}} + (\nu_\theta+\nu_\mu\,j_{1,n}^{\,2})\,a_n =\alpha_n(\hat{r}_s)\,L \,b_f\,Y,\\[0.5ex]
\frac{d b_n}{d\hat{t}} + (\nu_\phi+\nu_\mu\,j_{0,n}^{\,2})\,b_n =\epsilon\,\beta_n(\hat{r}_s)\,L\,b_f\,Y,\\[0.5ex]
\hat{\omega} = \hat{\omega}_0-\sum_{n=1,N} a_n-\sum_{n=1,N} b_n,\label{east3}
\end{eqnarray}
where
\begin{equation}\label{e103a}
f(X,Y)= \left\{
\begin{array}{ccc} 2/{\cal I}&\phantom{==}&(X^{\,2}+Y^{\,2})^{\,1/4}< 1\\[0.5ex]
2/[{\cal I}\,(X^{\,2}+Y^{\,2})^{\,1/4}]&&(X^{\,2}+Y^{\,2})^{\,1/4}\geq 1
\end{array}\right..
\end{equation}
Here, a slight modification has been made to (\ref{e103}) and (\ref{e104}) in the linear regime, $\hat{W}<1$, (in which case they are not valid anyway) in order
to render them non-singular at the origin of the $X$-$Y$ plane. 

\section{Results}
\subsection{Introduction}\label{sres1}
According to \cite{hu}, the density pump-out in DIII-D discharge \#158115 is due to mode penetration 
at the $m=-11$/$n=2$ rational surface, which lies at the bottom of the pedestal. The formation of a magnetic island chain
at the -11/2 surface leads to a local flattening of the plasma temperature and density profiles via parallel transport along magnetic field-lines.
In order to flatten the profiles, the island width must exceed  certain critical values that depend on the ratios of the relevant parallel and
perpendicular diffusivities at the rational surface \cite{hel}. For the case of the electron temperature, making use of the analysis of \cite{hel},
as well as the data in Table~\ref{table1}, we estimate the critical width to be
\begin{equation} \label{ee103}
  W_{{\rm crit}\,T_e} \simeq \left(\frac{\chi_\perp}{v_e\,r_s}\,\frac{1}{\epsilon_s\,s\,n}\right)^{1/3}r_s\simeq 7.9\times 10^{-3}\,{\rm m},
\end{equation}
where $v_e=\sqrt{T_e/m_e}$, $\epsilon_s=r_s/R_0$, $\chi_\perp$ is the perpendicular diffusivity, and $m_e$ the electron mass.  For the case of the density, we estimate the critical width to be
\begin{equation}\label{ee104}
  W_{{\rm crit}\,n_e} \simeq \left(\frac{\chi_\perp}{v_i\,r_s}\,\frac{1}{\epsilon_s\,s\,n}\right)^{1/3}r_s\simeq 3.1\times 10^{-2}\,{\rm m},
\end{equation}
where $v_i=\sqrt{T_i/m_i}$.
If both profiles are flattened  across the island chain then, as a consequence of the fact that the
equilibrium density gradient at the -11/2 surface greatly exceeds the temperature gradient (i.e., because $n_e/(T_i\,\eta_i)\simeq 3.5 \,
\,[10^{\,19}\,{\rm m}^{-3}/({\rm keV})]$; see Table~\ref{table1}, as well as  Figure~2 of \cite{hu}), the flattening naturally produces 
a much larger reduction in the pedestal density (measured in units of  $10^{\,19}\,{\rm m}^{-3}$) than in the pedestal temperature
(measured in units of keV).

According to \cite{hu},
ELM suppression in DIII-D discharge \#158115 is due to mode penetration at the $m=-8$/$n=2$ rational surface, which lies at the top of
the pedestal. The formation of a magnetic island chain at the -8/2 surface leads to a local flattening of the plasma temperature and density profiles.
As before, the island width must exceed certain critical values to flatten the profiles. We estimate the critical
width required to flatten the electron temperature profile to be [cf.\ (\ref{ee103}) and Table~\ref{table1}]
\begin{equation}
W_{{\rm crit}\,n_e}   \simeq 8.5\times 10^{-3}\,{\rm m},
\end{equation}
whereas the critical width required to flatten the density profile is [cf.\ (\ref{ee104}) and Table~\ref{table1}]
\begin{equation}
  W_{{\rm crit}\,n_e} \simeq 3.3\times 10^{-2}\,{\rm m}.
  \end{equation}
 The  flattening of the temperature and density profiles at the -8/2 rational surface is presumed  to be sufficient to prevent the plasma in the pedestal from ever exceeding the peeling-ballooning stability threshold, which leads  to ELM suppression. 

\subsection{Linear Simulations}\label{slin}
Table~\ref{table1} shows measured and estimated physics parameters at the $m=-8$/$n=2$ and $m=-11$/$n=2$ rational
surfaces in DIII-D discharge \#158115. Incidentally, all experimental minor radii quoted in this paper are flux-surfaced-averaged
minor radii, rather than minor radii on the outboard mid-plane. Making use of the analysis contained in the Appendices, these parameters
can be used to derive the input parameters for the linear model [i.e., (\ref{ast1})--(\ref{ast2})] that are listed in Table~\ref{table2}. 
Note that the dimensionless toroidal flow-damping rate has been set to zero (mostly because there is insufficient data to calculate its value). 
Note, further, that our knowledge of the edge current profile in DIII-D discharge \#158115 is insufficient to allow us to
calculate the parameters $\hat{\mit\Delta}'$ and ${\cal A}$ with any degree of accuracy. Fortunately, the penetration threshold
exhibits virtually no dependence on the parameter $\hat{\mit\Delta}'$, which can, therefore, safely be given its vacuum value of unity. With less justification, the amplification
parameter, ${\cal A}$, is also given its vacuum value unity.  
In DIII-D
discharge \#158115, the relative phase of RMPs generated by two sets of external field coils is  modulated sinusoidally at
a frequency of 1 Hz. This causes the amplitudes of the resonant harmonics of the applied RMP to modulated in a cycloidal manner at the same
freqeuncy. 

\begin{figure}[p]
\includegraphics[height=7in]{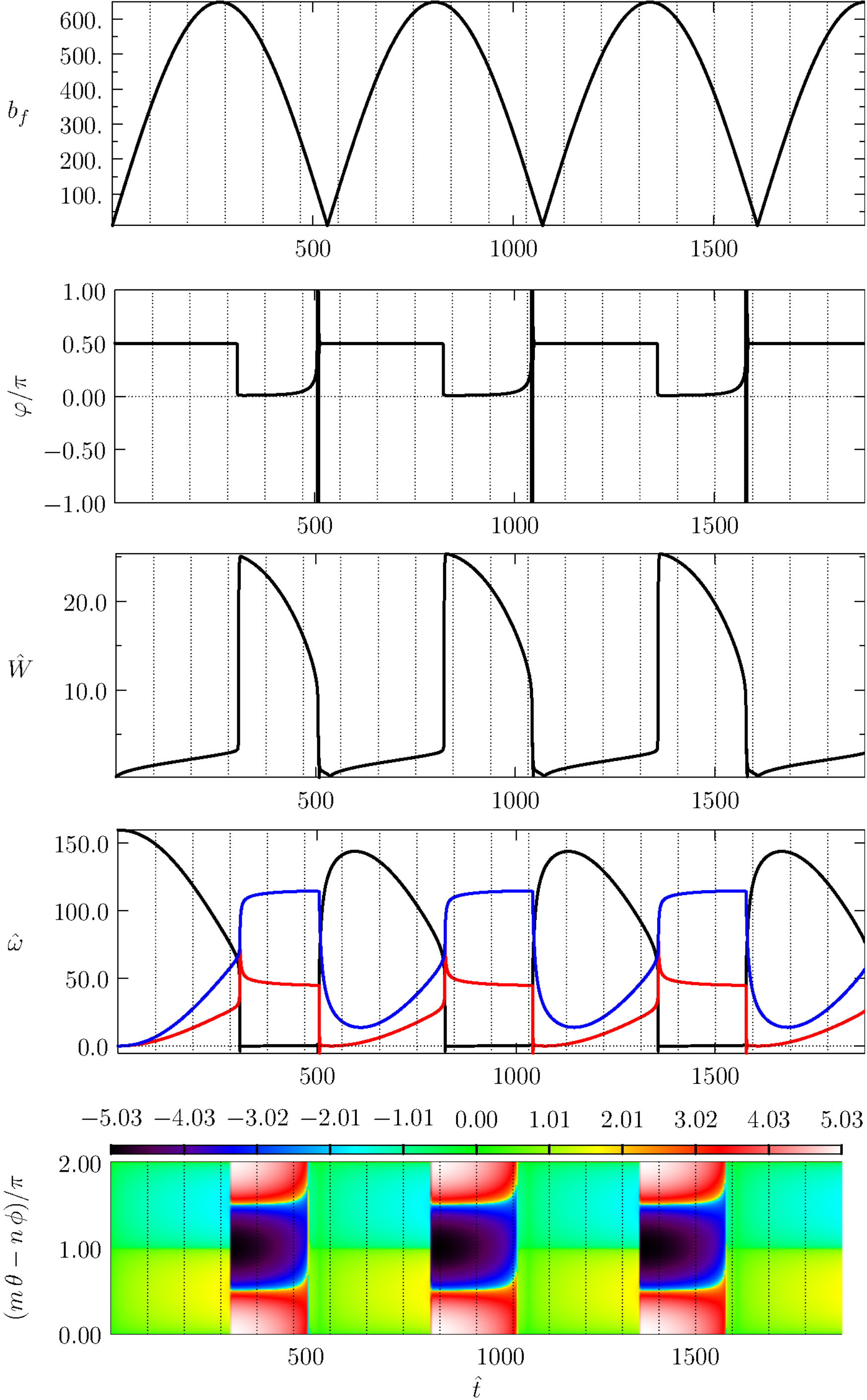}
\caption{Linear simulation of plasma response to an applied RMP at the $m=-8$/$n=2$ rational surface in DIII-D discharge
\#158115. The top panel shows the applied RMP. The second panel  shows the helical phase of the reconnected magnetic flux.
The third panel shows the RMP-induced magnetic island width (normalized to $\delta_{\rm SC}$). The fourth panel
shows the RMP-modified natural frequency, $\hat{\omega}$ (black curve), the  RMP-induced shift in the
plasma poloidal angular velocity, $\hat{\omega}_\theta$ (red curve), and the  RMP-induced shift in the
plasma toroidal angular velocity, $\hat{\omega}_\phi$ (blue curve). The previous three quantities are all
normalized to $1/\tau_{\rm SC}$.  The bottom panel shows simulated Mirnov data. To be more exact,
it shows contours of $\hat{W}^{\,2}\,\cos[(m\,\theta-n\,\phi-\varphi)]$. $\hat{t}$ is time normalized to $\tau_{\rm SC}$. } 
\label{fig1}
\end{figure}

\begin{figure}[p]
\includegraphics[height=7in]{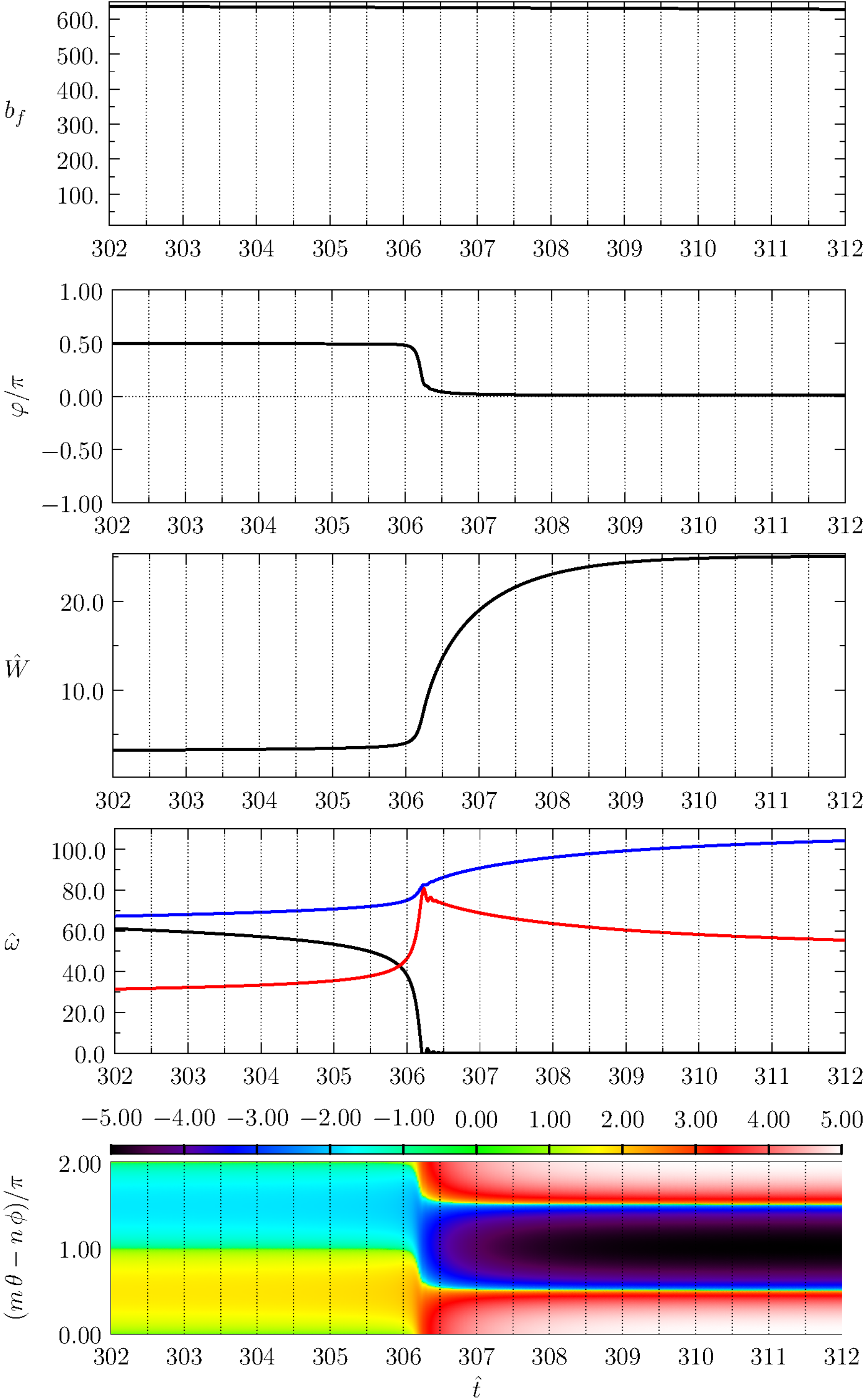}
\caption{Detail of Figure~\ref{fig1} showing mode penetration.  See Figure~\ref{fig1} caption.} \label{fig2}
\end{figure}

\begin{figure}[p]
\includegraphics[height=7in]{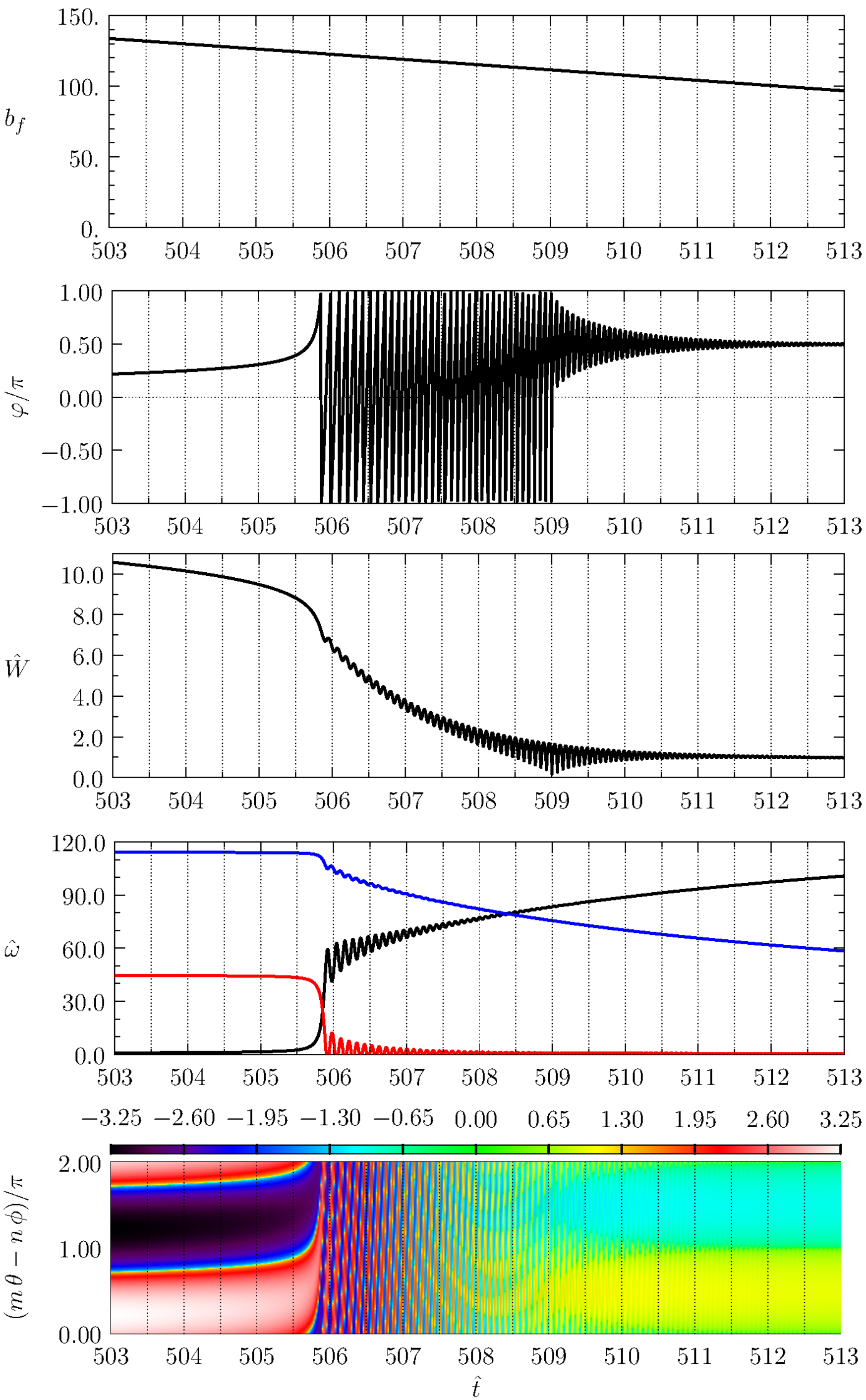}
\caption{Detail of Figure~\ref{fig1} showing mode unlocking.  See Figure~\ref{fig1} caption.} \label{fig3}
\end{figure}

Figure~\ref{fig1} shows a linear simulation of the response of the plasma at the -8/2 rational surface in
DIII-D discharge \#158115 to an RMP whose magnitude is switched on at $t=0$, and then modulated cycloidally at a frequency of 1 Hz. All simulations in this paper are performed with 200 velocity harmonics [i.e., $N=200$ in (\ref{ast2}) and (\ref{east3})].
The simulation data presented in Figure~\ref{fig1} is
qualitatively similar to the experimental data shown in Figure~1 of \cite{hu}, as well as the TM1 simulation
data displayed in Figure~2 of the same paper. In particular, if the amplitude of the applied RMP rises above a certain threshold
value then there is a bifurcation from a shielded solution characterized by $\hat{W}\ll \sqrt{b_f}$ to an unshielded solution characterized
by $\hat{W}\sim \sqrt{b_f}$. This bifurcation, which is known as {\em mode penetration}, is accompanied by a sudden reduction in
the natural frequency to zero, as well as sudden shifts in the plasma poloidal and toroidal rotation at the rational surface. 
Figure~\ref{fig2} shows the mode penetration process in more detail. Furthermore, referring again to Figure~\ref{fig1}, if the amplitude of the
applied RMP falls below a second smaller threshold value then there is a bifurcation from an unshielded solution to a
shielded solution. This bifurcation, which is known as {\em mode unlocking}, is accompanied by the recovery
of the natural frequency to its unperturbed value, as well   as sudden shifts in the plasma poloidal and toroidal rotation at the rational surface. Figure~\ref{fig3} shows the mode unlocking process in more detail. Note that, immediately after unlocking,
the magnetic island chain at the rational surface {\em spins-up}\/ and decays, before eventually re-locking to the RMP in a fixed helical phase relation. 

As is apparent from Figure~\ref{fig1}, mode penetration is triggered as soon as the natural frequency has been reduced to
approximately one half of its unperturbed value \cite{rf1}.
Moreover, about one third of the
reduction in the natural frequency associated with mode penetration is due to a shift in the local plasma poloidal rotation, whereas 
two thirds is due to a shift   in the local toroidal rotation. In the absence of neoclassical poloidal flow-damping, only about 2\% (i.e., a fraction $\epsilon$---see Table~\ref{table2}) of the
change in the natural frequency would be due to a shift in local plasma toroidal rotation. 
The large increase in the fraction of the overall frequency change due to  toroidal velocity shifts, in the presence of neoclassical poloidal flow-damping,   is a consequence of  the 
fact that the majority of charged particles in the DIII-D pedestal are trapped in banana orbits, and cannot,
therefore, freely rotate in the poloidal direction. Hence, neoclassical  poloidal flow-damping is comparatively strong in the pedestal
region of the DIII-D tokamak (compared to that in the plasma core). (The same is likely to be true in the pedestal regions of all relatively large,  conventional aspect-ratio
tokamaks.) The mode-penetration-induced shift  in the plasma toroidal rotation shown in Figure~\ref{fig1} corresponds to a toroidal
velocity shift in the co-current direction of about $30\,{\rm km/s}$ (see Table~\ref{table3}), which is in agreement with experimental
observations \cite{hu}.  The mode-penetration-induced shift  in the plasma poloidal rotation shown in Figure~\ref{fig1} corresponds to a poloidal 
velocity shift in the ion diamagnetic direction of only about $2\,{\rm km/s}$. Furthermore, we would expect the latter velocity
shift to be strongly localized in the vicinity of the rational surface \cite{cole}. Hence, it is not surprising that the poloidal
velocity shift is not observed experimentally \cite{paz1}.

According to Figures~\ref{fig2} and \ref{fig3}, the sudden collapse/recovery of the natural frequency associated with mode penetration/mode unlocking takes
place on a timescale of about a millisecond (see Table~\ref{table3}), and is accompanied by a sudden shift in the local plasma poloidal rotation
that takes place on the same timescale. The mode penetration/mode unlocking -induced shift in the local
plasma toroidal rotation takes place on a significantly longer timescale (at least, 10 milliseconds). 

There is one major difference between the simulation data shown in Figure~\ref{fig1} and the
experimental and TM1 simulation data shown in \cite{hu}. 
According to Figure~\ref{fig1},  mode penetration at the -8/2 rational surface in DIII-D discharge \#158115 occurs when $b_f$ exceeds the critical value $649$, which
corresponds to a vacuum radial field at the -8/2 surface of $b_v = 45$ gauss. (See Table~\ref{table3}.)
However, the penetration threshold inferred from experimental
data is more like $b_v=6$ gauss \cite{hu}.  This  discrepancy is probably related to  the fact that, according  to Figures~\ref{fig1} and \ref{fig2}, mode penetration at the -8/2
rational surface in DIII-D discharge \#158115  occurs when $W\sim 4\, \delta_{\rm SC}$.
In other words, when the magnetic island width exceeds the linear layer width. Given that the linear model
is only valid  when $W<\delta_{\rm SC}$, we conclude that mode penetration at the -8/2 rational surface in DIII-D discharge \#158115 is actually governed by nonlinear physics. This particular conclusion is not consistent with the TM1 simulations
described in \cite{hu}, according to which mode penetration at the -8/2 rational surface seems to be governed by
linear physics (because of the absence of island pulsations in the TM1 simulations---see Section~\ref{snon}).  One possible explanation for this disagreement is  that  plasma perpendicular
viscosity is artificially increased by a large factor in TM1 simulations, in order to mimic the effect of strong neoclassical poloidal flow-damping, but such an increase may also
artificially increase the linear layer width. Note, from Table~\ref{table3}, that the semi-collisional layer width
at the -8/2 rational surface is only $4\,{\rm mm}$, which is similar to the ion sound radius. On the other hand, the critical island
width above which mode penetration is triggered is about $1.6\,{\rm cm}$. Incidentally, after mode penetration has occurred, the
-8/2 island width rises well above the values required to locally flatten the temperature and density
profiles. (See Section~\ref{sres1}.) 

\begin{figure}[p]
\includegraphics[height=7in]{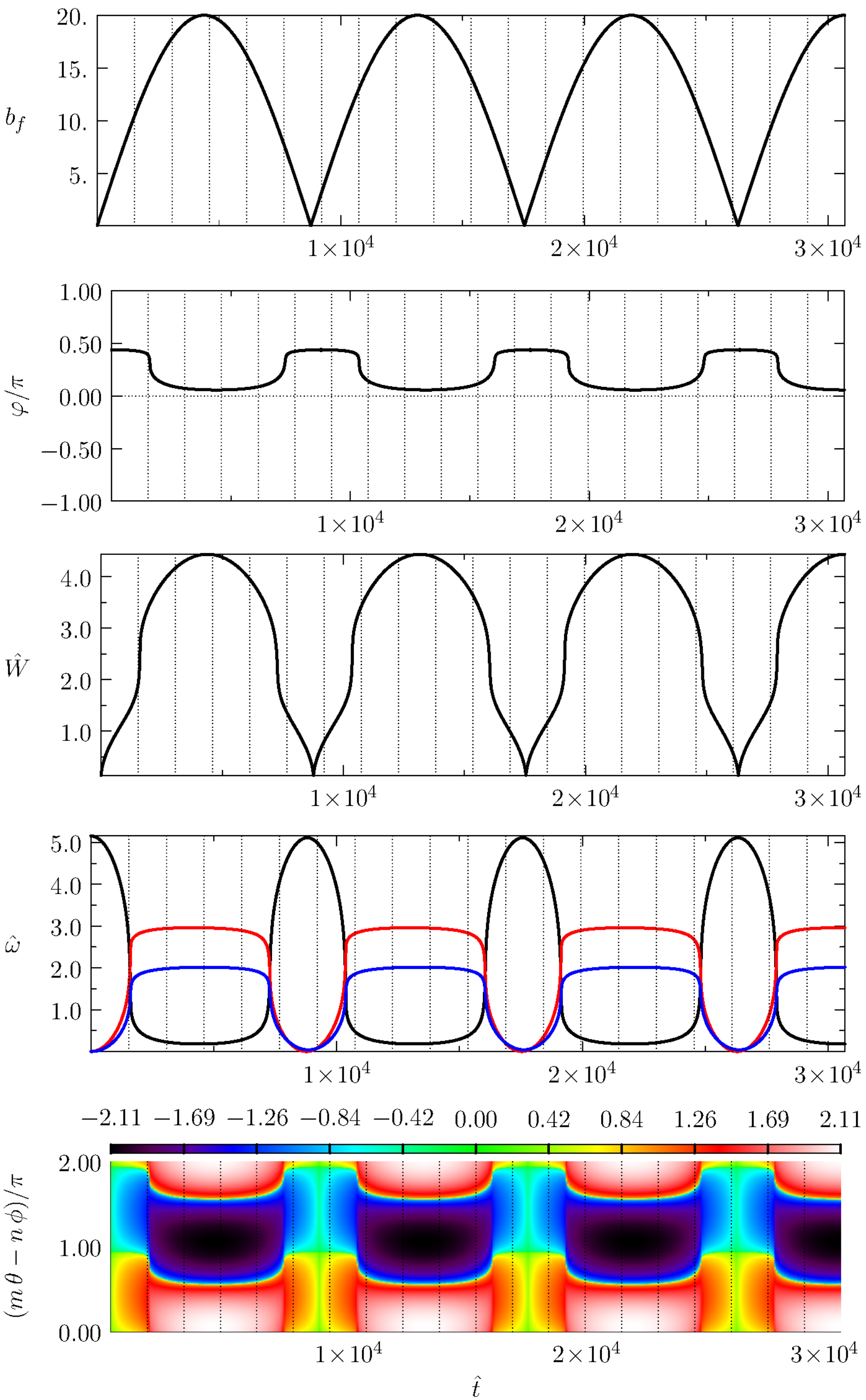}
\caption{Linear simulation of plasma response to an applied RMP at the $m=-11$/$n=2$ rational surface in DIII-D discharge
\#158115. See Figure~\ref{fig1} caption.} 
\label{fig4}
\end{figure}

Figure~\ref{fig4}   shows a linear simulation of the response of the plasma at the -11/2 rational surface in
DIII-D discharge \#158115 to an RMP whose magnitude is switched on at $t=0$, and then modulated cycloidally at a frequency of 1 Hz. 
The simulation data presented in Figure~\ref{fig4} is
qualitatively similar to  the TM1 simulation
data shown in Figure~2 of \cite{hu}. In particular, it  is clear from Figure~\ref{fig4} that there is insufficient  plasma
rotation at the -11/2 rational surface to enable the effective shielding of driven magnetic reconnection. In other words, $\hat{W}\sim \sqrt{b_f}$ at all times. The peak -11/2 island width is sufficiently large to locally flatten the temperature and density profiles. (See Section~\ref{sres1}.)
Note, however,  that $W>\delta_{\rm SC}$, which implies that driven reconnection
at the -11/2 rational surface in DIII-D discharge \#158115 is actually governed by nonlinear physics. 

\subsection{Nonlinear Simulations}\label{snon}
 Making use of the analysis contained in the Appendices, the experimental data given in Table~\ref{table1} 
can be used to derive the input parameters for the nonlinear model [i.e., (\ref{e103})--(\ref{e103a})] that are listed in Table~\ref{table2a}. 
Note that the dimensionless toroidal flow-damping rate has again been set to zero. 

\begin{figure}[p]
\includegraphics[height=7in]{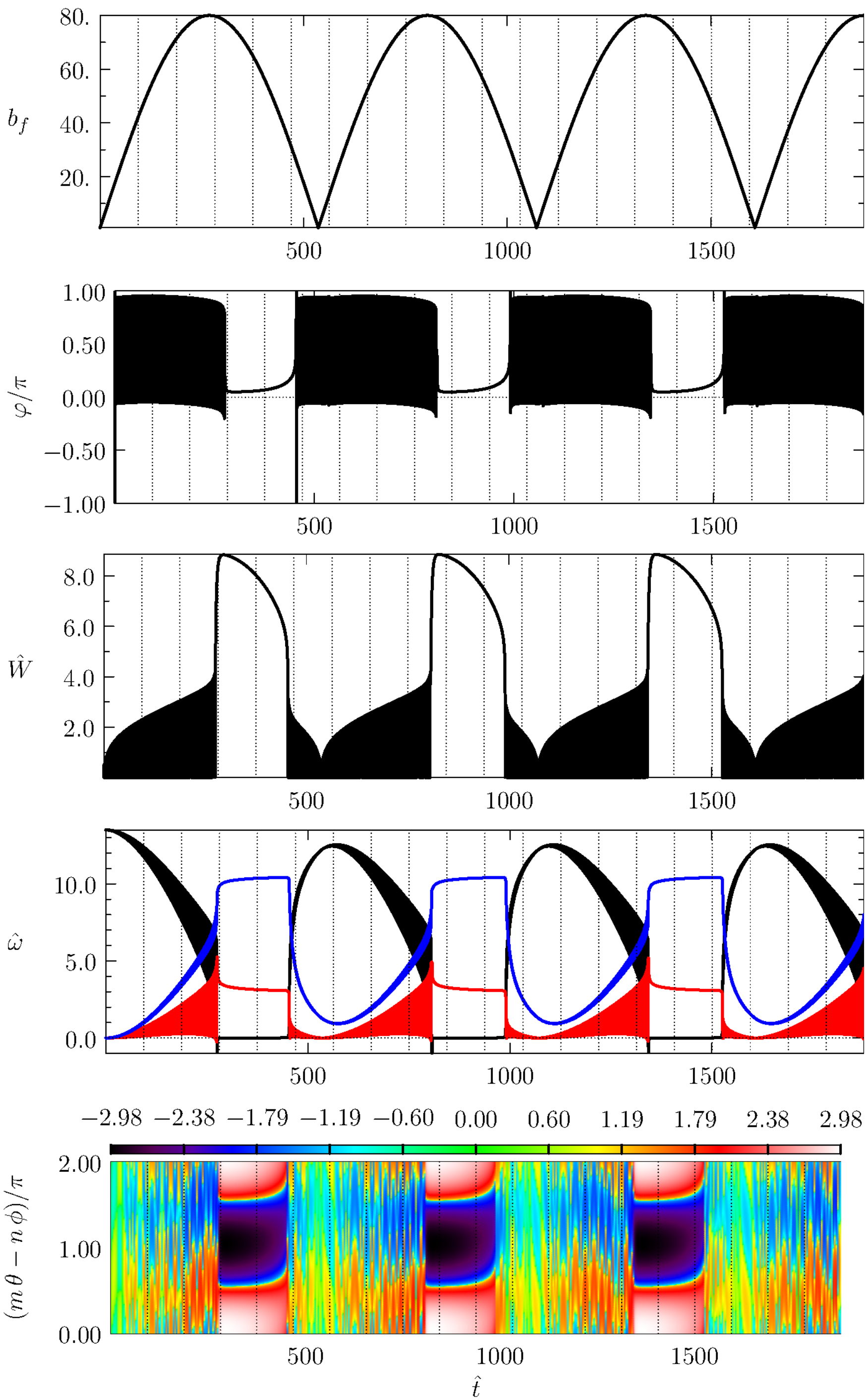}
\caption{Nonlinear simulation of plasma response to an applied RMP at the $m=-8$/$n=2$ rational surface in DIII-D discharge
\#158115. See Figure~\ref{fig1} caption.} 
\label{fig5}
\end{figure}

\begin{figure}[p]
\includegraphics[height=7in]{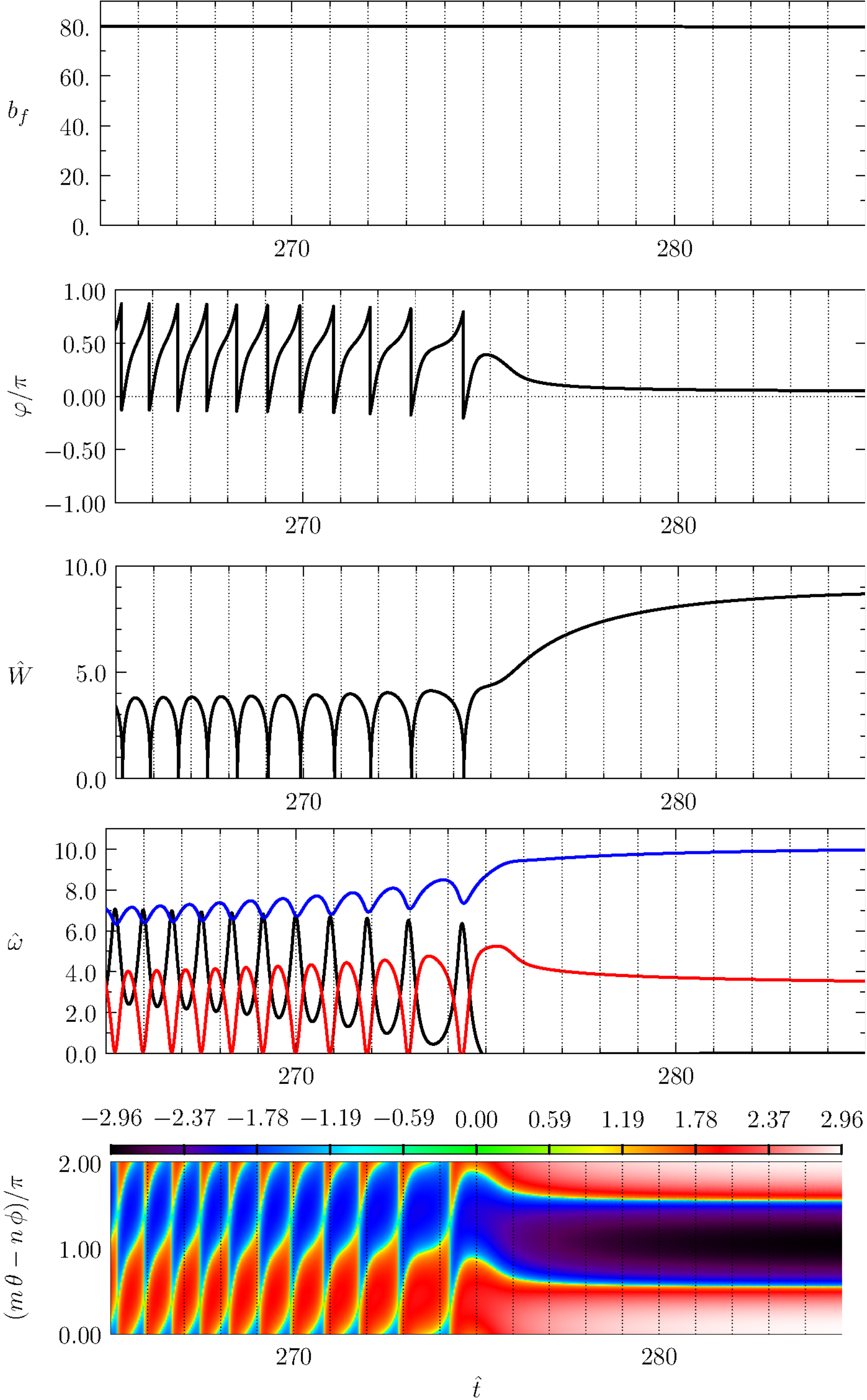}
\caption{Detail of Figure~\ref{fig5} showing mode penetration.  See Figure~\ref{fig1} caption.} \label{fig6}
\end{figure}

\begin{figure}[p]
\includegraphics[height=7in]{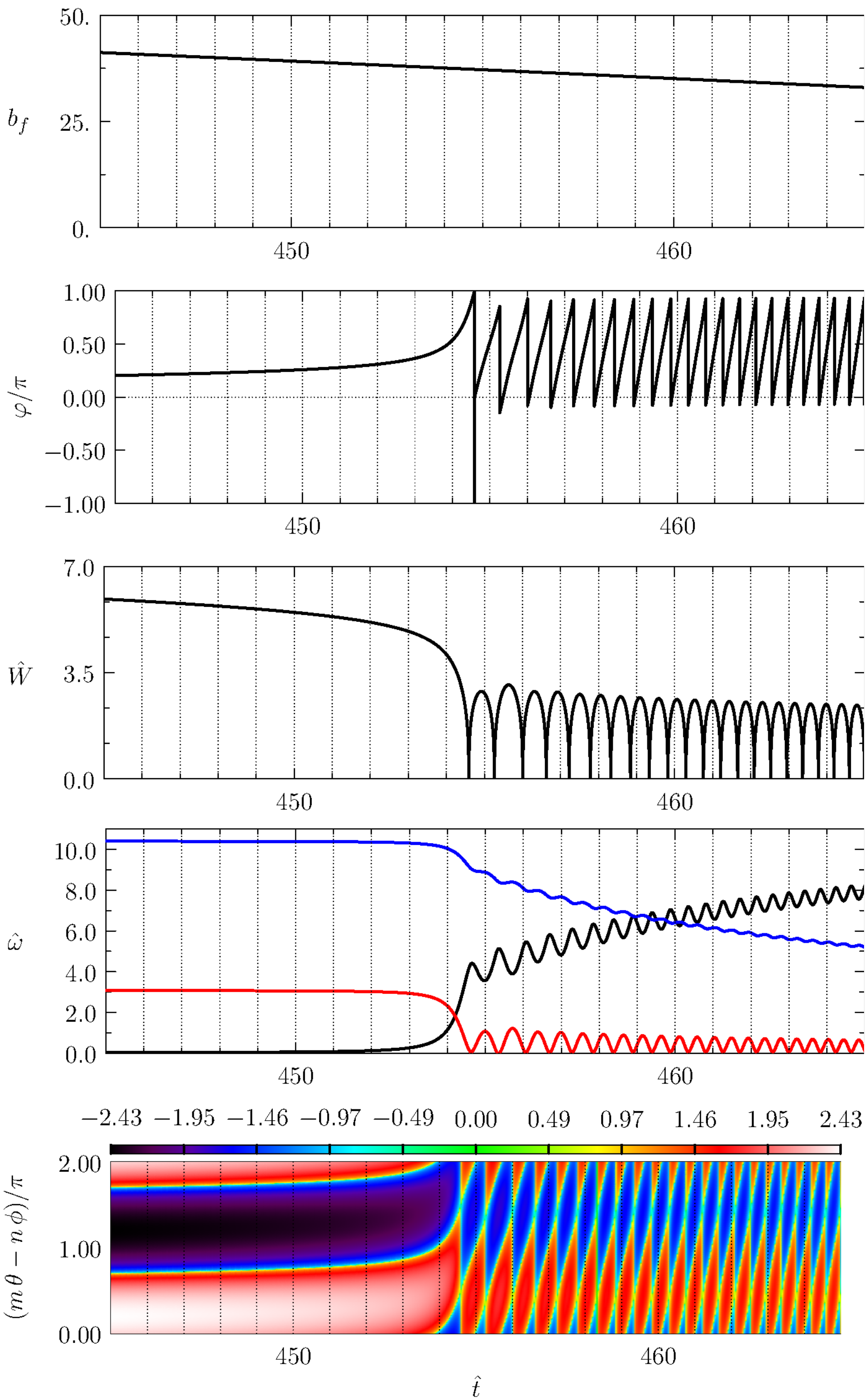}
\caption{Detail of Figure~\ref{fig5} showing mode unlocking.  See Figure~\ref{fig1} caption.} \label{fig7}
\end{figure}

Figure~\ref{fig5} shows a nonlinear simulation of the response of the plasma at the -8/2 rational surface in
DIII-D discharge \#158115 to an RMP whose magnitude is switched on at $t=0$, and then modulated cycloidally at a frequency of 1 Hz.
The simulation data presented in Figure~\ref{fig5} is 
qualitatively similar to the experimental data shown in Figure~1 of \cite{hu}. As before, if the amplitude of the applied RMP rises above a certain threshold value then mode penetration occurs. In other words, there is bifurcation from a shielded
to an unshielded solution, accompanied by a sudden reduction in
the natural frequency to zero, as well as  sudden shifts in the plasma toroidal and poloidal rotation at the rational surface. 
Figure~\ref{fig6} shows the mode penetration process in more detail. Furthermore, again referring to Figure~\ref{fig5}, if the amplitude of the
applied RMP falls below a second smaller threshold value then mode unlocking occurs. In other words, there is a bifurcation from an unshielded solution to a
shielded solution, accompanied by the recovery
of the natural frequency to its unperturbed value, as well   as  sudden shifts  in the plasma toroidal and poloidal rotation at the rational surface. Figure~\ref{fig7} shows the mode unlocking process in more detail.  
As in the linear case, mode penetration is triggered when the natural frequency has been reduced to about half of its
original value. Moreover, 
about two thirds of the change in the natural frequency associated with
mode penetration/mode unlocking is due to a shift in the local toroidal plasma rotation, and about one third to a
shift in the local poloidal plasma rotation. 

The main difference between the nonlinear simulations discussed  in this section and the linear simulations discussed in Section~\ref{slin}
lies in the nature of the shielded solution. In the linear simulations, the shielded solution consists of a narrow
magnetic island chain that has a fixed helical phase shift of about $+\pi/2$
 with respect to the locally resonant component of the RMP. On the other hand, in the nonlinear simulations, 
the shielded solution consists of a narrow island chain whose helical phase continually increases in time, and
whose width periodically falls to zero, at which times its helical
phase  jumps by $-\pi$ radians. This type of pulsating island solution was first predicted in \cite{rf2}, 
is discussed in detail in \cite{rfx, rfy, rff}, and has been observed experimentally (see Figure~29 in \cite{nazz}). Note, from Figures~\ref{fig5} and \ref{fig6}, that the width of the pulsating
island chain exceeds the linear layer width (i.e., $\hat{W}>1$) during most of its cycle, which justifies the nonlinear
approach employed in this section. 

Figure~\ref{fig4} implies that mode penetration at the -8/2 rational surface in DIII-D discharge \#158115 occurs when
$b_f$ exceeds the critical value $80$, which corresponds to a vacuum radial field at the -8/2 surface of
$b_v = 5.5$ gauss (see Table~\ref{table3}). This estimate for the penetration threshold is close to the experimentally inferred
value of 6 gauss \cite{hu}.

According to Section~\ref{nat}, the natural frequency in the linear regime is
\begin{equation}
\omega_0 = -n\,\omega_{\perp\,e},
\end{equation}
where $\omega_{\perp\,e}= \omega_E+ \omega_{\ast\,e}$. On the other hand, the natural frequency in the nonlinear
regime is 
\begin{equation}
\omega_0 = -n\,\omega_E - n\left(1-\frac{\eta_i\,\lambda_{\theta\,i}}{1+\eta_i}\right)\omega_{\ast\,i}.
\end{equation}
For the case of the -8/2 rational surface in DIII-D discharge \#158115, $\eta_i=1.9$ and $\lambda_{\theta\,i}=0.272$,
which implies that
\begin{equation}
\omega_0 =-n\left(\omega_E + 0.82\,\omega_{\ast\,i}\right),
\end{equation}
in the nonlinear regime. The fact that the natural rotation of a nonlinear magnetic island chain is offset in the ion diamagnetic
direction, rather than the electron diamagnetic direction, relative to the local ${\bf E}\times {\bf B}$ frame, leads to a much smaller prediction for the natural frequency in the nonlinear regime relative to that in the linear regime. This is the main reason why the predicted penetration threshold in the nonlinear regime is so much smaller than
that in the linear regime. 
 According to linear physics, we would expect
mode penetration at the -8/2 surface to be triggered when $\omega_{\perp\,e}$ passes through zero. Moreover,
once mode penetration has occurred, and the natural frequency becomes zero, we would expect $\omega_{\perp\,e}$
to be pinned to zero at the rational surface. On the other hand, according to nonlinear physics, we would expect mode penetration at the -8/2 surface to be triggered when some frequency offset from the ${\bf E}\times {\bf B}$ frequency  in the ion diamagnetic direction passes through zero. Moreover,
once mode penetration has occurred, and the natural frequency becomes zero, we would expect the offset frequency 
to be pinned to zero at the rational surface. In fact, experimental RMP-induced  ELM suppression data from the DIII-D tokamak is not consistent with $\omega_{\perp\,e}$ being the trigger frequency \cite{paz}, but rather some frequency
offset from $\omega_{\perp\,e}$ in the ion diamagnetic direction, which constitutes strong evidence that mode penetration at the top of the
pedestal in DIII-D RMP-induced  ELM suppression experiments is not governed by linear physics. Incidentally, there is
clear experimental evidence that the natural frequency of a nonlinear magnetic island chain is offset in the ion diamagnetic
direction relative to the local ${\bf E}\times {\bf B}$ frame \cite{omega1,omega2}.

\begin{figure}[p]
\includegraphics[height=7in]{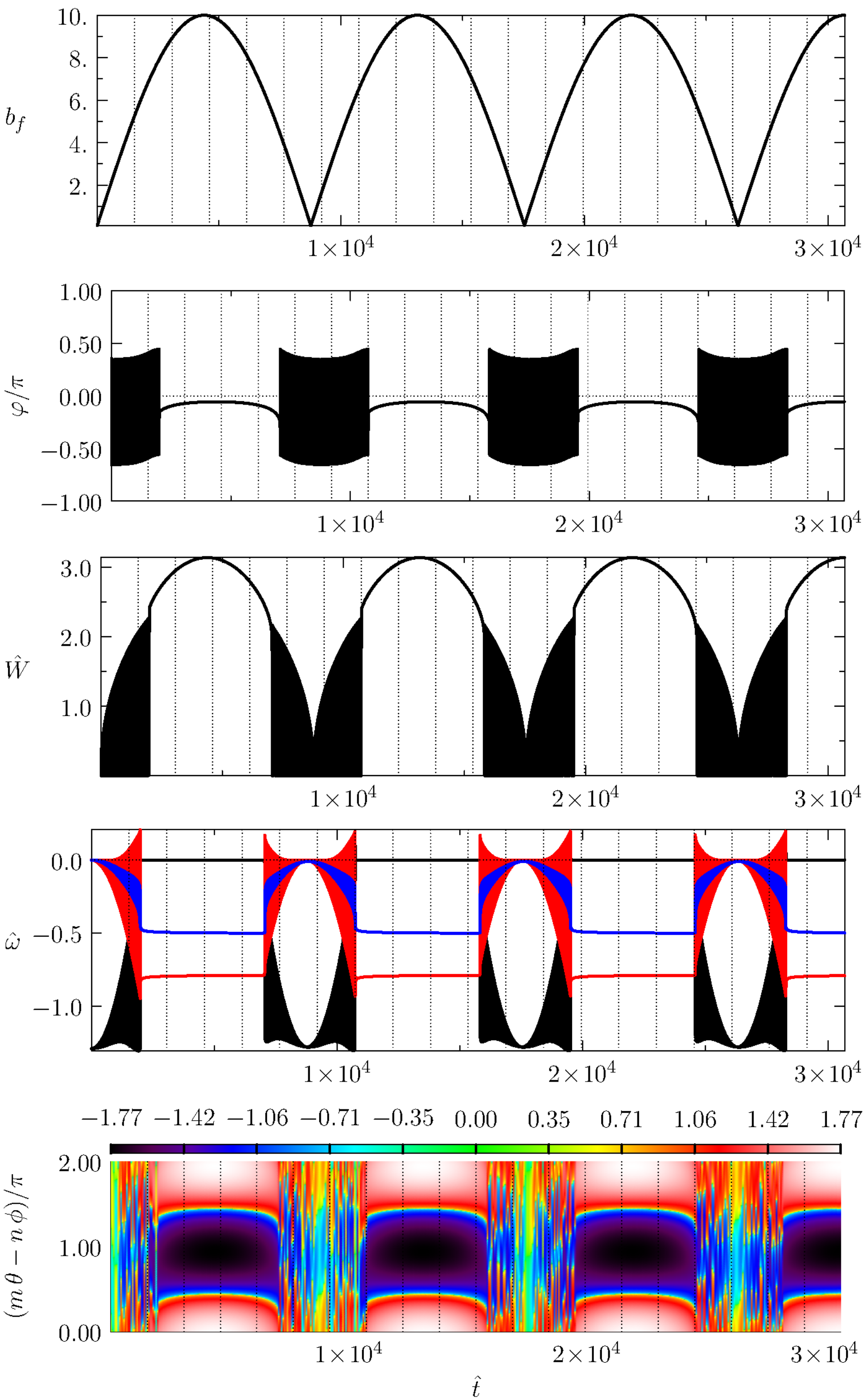}
\caption{Nonlinear simulation of plasma response to an applied RMP at the $m=-11$/$n=2$ rational surface in DIII-D discharge
\#158115. See Figure~\ref{fig1} caption.} 
\label{fig8}
\end{figure}

Figure~\ref{fig8}   shows a nonlinear simulation of the response of the plasma at the -11/2 rational surface in
DIII-D discharge \#158115 to an RMP whose magnitude is switched on at $t=0$, and then modulated cycloidally at a frequency of 1 Hz. 
The simulation data presented in Figure~\ref{fig8} is
qualitatively similar  the TM1 simulation
data shown in Figure~2 of \cite{hu}. In particular, it  is clear from Figure~\ref{fig8} that there is not enough plasma
rotation at the -11/2 rational surface to enable the effective shielding of driven magnetic reconnection. In other words, $\hat{W}\sim \sqrt{b_f}\equiv \hat{W}_f$ at all times. Note, however,  that $W>\delta_{\rm SC}$, which confirms that driven reconnection
at the -11/2 rational surface in DIII-D discharge \#158115 is  governed by nonlinear physics. 

\section{Summary and Discussion}
This paper investigates the plasma response to an externally generated, static, $n=2$, RMP,  whose amplitude is modulated cycloidally at
a frequency of 1 Hz, at two rational surfaces located  in the
pedestal of  DIII-D H-mode discharge \#158115 \cite{d158115}. The first rational surface is the $m=-8$/$n=2$ surface, and lies at the top
of the pedestal. The second is the $m=-11$/$n=2$ surface, and lies at the bottom of the pedestal. According to
the nonlinear  cylindrical reduced-MHD simulations of \cite{hu}, mode penetration at the -11/2 surface is responsible for the so-called density pump-out.
Moreover, mode penetration at the -8/2 surface is hypothesized to be responsible for RMP-induced ELM suppression. This
paper examines mode penetration at the -8/2 and -11/2 surfaces using two analytic plasma  response models. The
first model is {\em linear} in nature, and the second {\em nonlinear}. 

Our linear response model is based on the analysis of \cite{colef}, in which the four-field model \cite{four} is used to
find all possible  linear, two-fluid, drift-MHD, resonant plasma response regimes when a static RMP is applied to a large aspect-ratio tokamak plasma. We deduce that the particular response regime that is appropriate at both the -8/2 and -11/2 rational surfaces in DIII-D
discharge \#158115 is the so-called first semi-collisional regime (SCi). Incidentally, it has long been recognized that linear
(tearing) layer physics in high-temperature tokamak plasmas is semi-collisional in nature, rather than collisional or
collisionless \cite{drake,semi,semi1}. Our linear response model does not incorporate the screening effect due to magnetic 
curvature that was discovered in \cite{liu}. However, this effect may be negated  by parallel thermal transport \cite{bai}.
More importantly, the curvature screening effect is a prediction of collisional layer physics, and the true
layer physics in tokamak plasmas is semi-collisional. Indeed, \cite{semi} found that, in a semi-collisional
layer, the effect of the perturbed bootstrap current  is much greater in magnitude than, and opposed to, the effect of magnetic curvature. A much more serious deficiency in our linear layer model emanates from the fact that the layer width
is similar to the ion sound radius. Given that the ion and electron temperatures in the pedestal of DIII-D discharge \#158115
are almost equal, this implies that the layer width is also similar to  the  ion gyroradius. Unfortunately, the finite ion gyroradius width  is not taken into account in the layer analysis of \cite{colef}. Moreover, it is known that, in situations in which the ion gyroradius
is similar to, or exceeds, the linear layer width, finite ion orbit width effects can significantly modify the layer response \cite{semi,ion}.

The linear response model adopted in this paper is augmented by plasma poloidal and toroidal equations of motion that determine how the
quasi-linear electromagnetic locking torque that develops at the rational surface, in response to the
applied RMP, modifies the local plasma rotation. The equations of motion take plasma perpendicular viscosity, neoclassical
poloidal flow-damping, and neoclassical toroidal flow-damping into account. When the equations of motion are
combined with the linear response model, a closed set of equations is obtained; these equations are solved numerically. 

The linear response model is only valid when the width of the RMP-induced magnetic island chain at the
rational surface falls below the linear layer width. In the opposite situation, in which the driven island width
exceeds the linear layer width, the linear  response model must be replaced by a nonlinear response model. 
It turns out that the appropriate nonlinear response model is, in many ways, simpler than the linear response model,
given that it essentially consists of the Rutherford island width evolution equation \cite{ruth} combined
with the no-slip constraint \cite{rf1}.

The main  conclusion of our linear simulations, which are described in Section~\ref{slin}, is that a linear response model is
inapplicable at both the -8/2 and -11/2 rational surfaces in DIII-D discharge \#158115. The problem is that the semi-collisional
layer widths at the rational surfaces are so small (a few mm, in both cases)  that  rotational shielding is not sufficient to
reduce the driven magnetic island widths below the layer widths. 

Our nonlinear simulations, which are described in Section~\ref{snon}, give results that are  similar to the
experimental results described in \cite{hu}. At the -8/2 rational surface, driven magnetic reconnection
is strongly screened by plasma rotation as long as the resonant component of the radial magnetic field remains below a certain threshold value. However, as soon
as the radial field exceeds the threshold value, which is about 6 gauss, there is a sudden and irreversible breakdown of screening, accompanied by
rapid shifts in the local plasma toroidal and poloidal angular velocities. 
On the other hand, at the -11/2 rational surface, the plasma rotation is not large enough to screen driven magnetic reconnection. 

\section{Future Work}
There are a number
of improvements that could be made to the analytic model  described in this paper.  Such improvements include; using a more accurate neoclassical model; taking into account the  coupling
of different rational surfaces via mode-penetration-induced changes  in the plasma rotation, density, and temperature, profiles  in the pedestal \cite{hu};
taking into account the coupling of different poloidal harmonics of the RMP due to
toroidicity, the Shafranov shift, and flux surface shaping \cite{tor}; employing a more realistic plasma equilibrium; 
including island saturation terms \cite{sat1,sat2,sat3},
the perturbed bootstrap current \cite{hel},  and the perturbed ion
polarization current \cite{ionpz}, in the Rutherford equation; and taking into account orbit-squeezing effects due to the strong shear in the
radial electric field that is typically present in H-mode tokamak pedestals \cite{orbit}.

It is a well-known fact that RMP-induced ELM suppression is only effective when the value of $q_{95}$ (i.e., the safety-factor at the 95\% magnetic flux-surface)
lies in certain narrow windows \cite{d158115,pazs}.
In future work, we intend to use an improved version of the analytic model of RMP-induced ELM suppression presented in this paper to
investigate the dependence of the ELM suppression threshold on $q_{95}$.
Such an investigation will inevitably entail thousands of simulations, and is only feasible with the type of highly-reduced analytic model
described in this paper.

\section*{Acknowledgements}
This research was funded by the U.S.\ Department of Energy under contract DE-FG02-04ER-54742.
The author would like to thank R.~Nazikian, Q.M.~Hu, and C.~Paz-Soldan for helpful discussions.

\appendix

\section{Electron Neoclassical Effects}\label{appz}
It is helpful to define the  electron collisionality at the rational surface \cite{sauter}: 
\begin{equation}
\nu_{e\,\ast}= 1.13\times 10^{-3}\left[\frac{|m|}{n\,\hat{r}_s^{\,3/2}}\right]
\left[\frac{R_0^{\,5/2}\,n_e\,Z_{\rm eff}}{a^{\,3/2}\,T_e^{\,2}}\right].
\end{equation}
Here, $Z_{\rm eff}$ is the effective ion charge number  (incidentally, the majority ion charge number is unity), $n_e$ the equilibrium  electron number density at the rational surface,  and $T_e$ the equilibrium electron temperature
at the rational surface. Moreover, $R_0$ is measured in meters, $a$ in meters, 
$n_e$ in $10^{19}\,{\rm m}^{-3}$, and $T_e$ in kilo-electron-volts. In accordance with the analysis of \cite{grob},
the Coulomb logarithm is assumed to take the value 17 for all plasma species. 
Now, the fraction of trapped particles at the rational surface, assuming that the plasma
there lies in the banana collisionality regime,  is \cite{grob}
\begin{equation}
f_t=1.46\left[\hat{r}_s^{\,1/2}\right]\left[\frac{a^{\,1/2}}{R_0^{\,1/2}}\right] - 0.46\left[\hat{r}_s^{\,3/2}\right]\left[\frac{a^{\,3/2}}{R_0^{\,3/2}}\right].
\end{equation}
Let \cite{sauter,hirsh1,zarn}
\begin{equation}
X = \frac{f_t}{1+(0.55-0.1\,f_t)\,\nu_{e\,\ast}^{\,1/2}+0.45\,(1-f_t)\,\nu_{e\,\ast}/Z_{\rm eff}^{\,{3/2}}},
\end{equation}
and 
\begin{eqnarray}
F_e&=& \frac{Z_{\rm eff}}{1-(1 + 0.36/Z_{\rm eff})\,X+(0.59/Z_{\rm eff})\,X^{\,2}-(0.23/Z_{\rm eff})\,X^{\,3}}\nonumber\\[0.5ex]&\phantom{=}&\times\frac{1+1.198\,Z_{\rm eff} + 0.222\,Z^{\,2}_{\rm eff}}{1+2.966\,Z_{\rm eff} + 0.753\,Z_{\rm eff}^{\,2}}.
\end{eqnarray}
We can define the {\em effective electron temperature}\/ at the rational surface:
\begin{equation}
T_{e\,{\rm eff}} = \frac{T_e}{F_e^{\,2/3}}.
\end{equation}
This quantity is the electron temperature that gives the correct plasma resistivity, taking into
account the effect of impurities and the neoclassical modification of plasma resistivity \cite{sauter,hirsh1}, when plugged into the standard formula  $\eta_\parallel = m_e/(n_e\,e^{\,2}\,\tau_{ee})$. Here, $\tau_{ee}$ is electron/electron
$90^\circ$ scattering timescale at the rational surface. 

\section{Ion Neoclassical Effects}\label{appz1}
It is helpful to define the  ion collisionality at the rational surface \cite{sauter}: 
\begin{equation}
\nu_{i\,\ast}= 9.07\times 10^{-4}\left[\frac{|m|}{n\,\hat{r}_s^{\,3/2}}\right]
\left[\frac{R_0^{\,5/2}\,n_e}{a^{\,3/2}\,T_i^{\,2}}\right].
\end{equation}
Here, $T_i$ is the ion temperature at the rational surface, measured in kilo-electron-volts.  
Let \cite{sigmar,grob,grob1,callen}
\begin{eqnarray}
\hat{K}_{00\,{\rm B}}&=& \alpha+0.533,\\[0.5ex]
\hat{K}_{00\,{\rm P}} &=& 1.77,\\[0.5ex]
\hat{K}_{00\,{\rm PS}}&=& \frac{4.25\,\alpha +3.02}{D},\\[0.5ex]
\hat{K}_{01\,{\rm B}}&= &\alpha+0.707,\\[0.5ex]
\hat{K}_{01\,{\rm P}} &=& 5.32,\\[0.5ex]
\hat{K}_{01\,{\rm PS}}&= &\frac{20.13\,\alpha +12.43}{D},\\[0.5ex]
\hat{K}_{11\,{\rm B}}&= &2\,\alpha+1.591,\\[0.5ex]
\hat{K}_{11\,{\rm P}} &=& 21.27,\\[0.5ex]
\hat{K}_{11\,{\rm PS}}&=& \frac{101.06\,\alpha +58.65}{D},\\[0.5ex]
D &= &2.40\,\alpha^{\,2}+5.32\,\alpha+2.225,\\[0.5ex]
\alpha &= &Z_I\left(\frac{Z_{\rm eff}-1}{Z_I-Z_{\rm eff}}\right).
\end{eqnarray}
Here, $Z_i=1$ and $Z_I=6$ are the charge numbers of the majority (${\rm H}^{\,2}$) and impurity (${\rm C}^{\,6+}$) ions, respectively. Note that we
are making the simplifying assumption  that the impurity ion mass is much larger than the majority ion mass, that the impurity ion neoclassical
viscous force is negligible compared to the friction force acting between the two ion species, and that the two ion species have the same temperature \cite{grob}. 
It follows that \cite{grob,sigmar,grob1,callen}
\begin{eqnarray}
\fl \hat{K}_{ab}= \frac{\hat{K}_{ab\,{\rm B}}}
{(1+2.92\,\nu_{i\,\ast}\,\hat{K}_{ab\,{\rm B}}/\hat{K}_{ab\,{\rm P}})
\,[1+2\,\epsilon_s^{\,3/2}\,\nu_{i\,\ast}\,\hat{K}_{ab\,{\rm P}}/(3\,\hat{K}_{ab\,{\rm PS}})]},
\end{eqnarray}
for $a$, $b=0,1$. The normalized ion neoclassical viscosities are written
\begin{eqnarray}
\hat{\mu}_{00\,i} &=& g\,\hat{K}_{00},\\[0.5ex]
\hat{\mu}_{01\,i} &=&g\left( \frac{5}{2}\,\hat{K}_{00} -\hat{K}_{01}\right),\\[0.5ex]
\hat{\mu}_{11\,i}&=& g\left(\hat{K}_{11} -5\,\hat{K}_{01} +\frac{25}{4}\,\hat{K}_{00}\right),
\end{eqnarray}
where $g=f_t/(1-f_t)$. 
 
Let
\begin{equation}
F_i = \left(\frac{q_s}{\epsilon_s}\right)^2\,\hat{\mu}_{00\,i}.
\end{equation}
The neoclassical poloidal flow-damping timescale takes the form \cite{grob,hirsh1}
\begin{equation}
\tau_\theta = \frac{\tau_{ii}}{F_i},
\end{equation}
where $\tau_{ii}$ is the majority-ion/majority-ion $90^\circ$ scattering timescale at the rational surface \cite{plasma}. It is helpful to define the {\em effective ion temperature}\/ at the rational surface:
\begin{equation}
T_{i\,{\rm eff}} = \frac{T_i}{F_i^{\,2/3}}.
\end{equation}
This is the temperature at which the neoclassical poloidal flow-damping timescale matches the majority-ion/majority-ion $90^\circ$ scattering timescale at the rational surface. 

According to \cite{rff}, the natural frequency of a nonlinear magnetic island chain
takes the form
\begin{equation}\label{a14}
\omega_0=-n\,\omega_E-n\left(1-\frac{\eta_i\,\lambda_{\theta\,i}}{1+\eta_i}\right)\omega_{\ast\,i},
\end{equation}
where \cite{grob,sigmar,grob1,callen}
\begin{equation}\label{a15}
\lambda_{\theta\,i} =\frac{\hat{\mu}_{01\,i}}{\hat{\mu}_{00\,i}+(\hat{\mu}_{00\,i}\,\hat{\mu}_{11\,i}-\hat{\mu}_{01\,i}^{\,2})/(\sqrt{2}+\alpha-\alpha\,\beta)},
\end{equation}
and
\begin{equation}\label{a16}
\beta = \left.\left(\frac{27}{4}\right)^2\left(\frac{M_i}{M_I}\right)^2\right/\left(\frac{15}{2} +\sqrt{\frac{2\,\alpha\,M_I}{M_i}}\right).
\end{equation}
Here, $M_i=2$ and $M_I=12$ are the mass numbers of the majority and minority ions, respectively. 

\section{Model Parameters}\label{appa}
The hydromagnetic timescale is written
\begin{equation}
\tau_H= 1.45\times 10^{-7}\left[\frac{M_i^{1/2}}{n\,s}\right]\left[\frac{R_0\,n_e^{1/2}}{|B_\phi|}\right],
\end{equation}
where $\tau_H$ is measured in seconds,  and
$B_\phi$ in tesla.
The resistive timescale is written
\begin{equation}
\tau_R= 2.27\times 10^{+1}\left[\hat{r}_s^{\,2}\right]\left[a^{\,2}\,T_{e\,{\rm eff}}^{\,3/2}\right],
\end{equation}
where $\tau_R$ is measured in seconds.
The viscous diffusion timescale is written
\begin{equation}
\tau_V = \left[\hat{r}_s^{\,2}\right]\left[\frac{a^{\,2}}{\chi_\perp}\right],
\end{equation}
where $\tau_V$ is measured in seconds, and the perpendicular momentum diffusivity at the rational surface, $\chi_\perp$, is measured in 
meters squared per second. 

The linear layer width is written
\begin{equation}
\delta_{\rm SC} = 9.36\times 10^{-4}\left[\frac{\hat{r}_s}{n\,s}\right]\left[\frac{R_0\,a\,n_e^{1/2}\,|n\,\omega_{\ast\,e}|^{\,1/2}}{T_e^{\,1/2}\,T_{e\,{\rm eff}}^{\,3/4}}\right],
\end{equation}
where $\delta_{\rm SC}$ is measured in meters. 
The linear reconnection time is written
\begin{equation}
\tau_{\rm SC} = 1.06\times 10^{-2}\left[\frac{\hat{r}_s^{\,2}}{n\,s\,|m|}\right]\left[\frac{R_0\,a^{\,2}\,n_e^{1/2}\,T_{e\,{\rm eff}}^{\,3/4}\,|n\,\omega_{\ast\,e}|^{\,1/2}}{T_e^{\,1/2}}\right],
\end{equation}
where $\tau_{\rm SC}$ is measured in seconds. 

The dimensionless plasma viscosity parameter is written
\begin{equation}
\nu_\mu = 1.06\times 10^{-2}\left[\frac{\hat{r}_s^{\,2}}{n\,s\,|m|}\right]\left[\frac{R_0\,n_e^{1/2}\,T_{e\,{\rm eff}}^{\,3/4}\,\chi_\perp\,|n\,\omega_{\ast\,e}|^{\,1/2}}{T_e^{\,1/2}}\right].
\end{equation}
The dimensionless plasma poloidal flow-damping parameter is written
\begin{equation}
\nu_\theta =2.74\times 10^{+0} \left[\frac{\hat{r}_s^{\,2}}{M_i^{1/2}\,n\,s\,|m|}\right]\left[\frac{R_0\,a^{\,2}\,n_e^{3/2}\,T_{e\,{\rm eff}}^{\,3/4}\,|n\,\omega_{\ast\,e}|^{\,1/2}}{T_e^{\,1/2}\,T_{i\,{\rm eff}}^{\,3/2}}\right].
\end{equation}
The dimensionless locking parameter, $L$, is written
\begin{equation}
L= 1.61\times 10^{-5}\left[\frac{\hat{r}_s^{\,8}}{M_i\,n^{\,4}\,s^{\,4}\,|m|}\right]\left[
\frac{R_0^{\,4}\,a^{\,4}\,n_e^{\,2}\,B_\phi^{\,2}\,|n\,\omega_{\ast\,e}|^{\,3}}{T_e^{\,3/2}\,\,T_{e\,{\rm eff}}^{\,3/2}}
\right].
\end{equation}

The normalized natural frequency is written
\begin{equation}
\hat{\omega}_0= 1.06\times 10^{+1}\left[\frac{\hat{r}_s^{\,2}}{n\,s\,|m|}\right]\left[\frac{R_0\,a^{\,2}\,n_e^{1/2}\,T_{e\,{\rm eff}}^{\,3/4}\,|n\,\omega_{\ast\,e}|^{\,1/2}}{T_e^{\,1/2}}\right]\omega_0,
\end{equation}
where $\omega_0$ is measured in kilo-radians per second. 
The normalized radial magnetic field is written
\begin{equation}
b_f = 1.83\times 10^{+3}\left[\frac{n\,s}{\hat{r}_s}\right]\left[\frac{T_e\,T_{e\,{\rm eff}}^{\,3/2}}{R_0\,a\,n_e\,|B_\phi|\,|n\,\omega_{\ast\,e}|}\right]{\cal A}\,b_v,
\end{equation}
where the vacuum radial magnetic field at the rational surface, $b_v$, is measured in gauss. 

\begin{table}[p]
\begin{tabular}{cccccccccc}\hline
$m$~ & ~$n$~ & $B_\phi$~ & ~$R_0$~ & ~$a$~ ~ & ~$n_e$~ & ~$T_e$~ & ~$T_i$~ &~$\eta_i$~ & ~$Z_{\rm eff}$ \\[0.5ex]\hline
-8 & $2$ &$-1.94$ & 1.75 & 0.93 & 2.8 & 1.4 & 1.4  & 1.9& 2.5\\[0.5ex]
-11 & $2$ &$-1.94$ & 1.75 & 0.93 & 0.75 & 0.12 & 0.12  & 1.8 & 2.5 \\[0.5ex]\hline
$m$~ & ~$n$~ &   ~$\chi_\perp$~ & ~$\omega_E$~ &~$\omega_{\ast\,e}$~ & ~$\hat{r}_s$~ & ~$s$~ & ~$M_i$~ & ~${\mit\Delta}$~ &~${\cal A}$\\[0.5ex]\hline
-8 & $2$ & 1.0 & $-21.3$& $-21.5$ & 0.853 & 2.3 & 2.0 & 1.0 & 1.0\\[0.5ex]
 -11& $2$& 1.0 & $-9.8$& $-12.8$ & 0.974 & 11.2 & 2.0 & 1.0& 1.0\\[0.5ex]\hline
\end{tabular}
\caption{Measured and estimated physics parameters at two rational surfaces in the pedestal of DIII-D discharge
\#158115 (see Figure~2 of \cite{hu}).  $m$ is the poloidal mode number,
$n$ the toroidal mode number, $B_\phi$  the toroidal magnetic field (T), $R_0$ the major radius (m),  $a$ the minor radius (m),
$n_e$ the electron number density ($10^{\,19}\,{\rm m}^{-3}$), $T_e$ the electron temperature (keV), $T_i$
the ion temperature (keV), $\eta_i=d\ln T_i/d\ln n_e$, $Z_{\rm eff}$ the conventional measure of impurity content, $\chi_\perp$
the perpendicular momentum diffusivity (${\rm m}^{\,2}\,{\rm s}^{-1}$), $\omega_E = E_r/(R_0\,B_\theta)$ the ${\bf E}\times {\bf B}$ frequency
(${\rm krad}\,{\rm s}^{-1}$), $\omega_{\ast\,e}=(dp_e/dr)/(e\,n_e\,R_0\,B_\theta)$ the electron diamagnetic frequency  (${\rm krad}\,{\rm s}^{-1}$), $\hat{r}_s$ the rational surface radius normalized to the plasma minor radius, $s$ the magnetic shear,
$M_i$ the majority ion mass number, ${\mit\Delta}\equiv {\mit\Delta}'\,r_s/(2\,m)$, and ${\cal A}$ is the amplification factor. }\label{table1}
\end{table}

\begin{table}[p]
\begin{tabular}{cccccc}\hline
$m$~ & ~$n$~ & ~$c_\beta$~ & ~$D$~ & ~$P$~ &  ~$Q$~\\[0.5ex]\hline
-8 & $2$ &~$6.48\times 10^{-2}$ &$1.54\times 10^{+0}$& ~$7.16\times 10^{+0}$ & $1.27\times 10^{+0}$\\[0.5ex]
-11 & $2$ &~$9.81\times 10^{-3}$&$3.43\times 10^{-1}$ &~$4.63\times 10^{-1}$ &~$6.16\times 10^{-2}$
\end{tabular}
\caption{Input parameters for the analytic, cylindrical, single-helicity, four-field, linear, resonant plasma response model of \cite{colef} at two rational surfaces in the pedestal of DIII-D discharge
\#158115. $m$ is the poloidal mode number,
$n$ the toroidal mode number, $c_\beta = \sqrt{\beta}$ (where $\beta$ is the usual dimensionless measure of plasma pressure),
$D=S^{\,1/3}\,\rho_s/r_s$ (where $S=\tau_R/\tau_H$), $P=\tau_R/\tau_V$, and $Q=S^{\,1/3}\,|\omega_0|\,\tau_H/2$, with 
$\omega_0=-n\,(\omega_E+\omega_{\ast\,e})$. }\label{table1a}
\end{table}

\begin{table}[p]
\begin{tabular}{ccccccccc}\hline
$m$~ & ~$n$~ & ~$\hat{r}_s$~& ~$\epsilon$~& ~$\nu_\theta$~  &~$\nu_\phi$~ & 
$\nu_\mu$ & $L$ & $\hat{\omega}_0$ \\[0.5ex]\hline
-8 & $2$ & ~$0.853$ & ~$1.77\times 10^{-2}$~ & $4.35\times 10^{+2}$~ & $0.0$ &  $2.16\times 10^{-3}$~ & $9.41\times 10^{-3}$~ & $1.60\times 10^{+2}$\\[0.5ex]
-11 & $2$ & ~$0.974$ & ~$9.34\times 10^{-3}$~ & $6.50\times 10^{+1}$~ & $0.0$ & $1.32\times 10^{-4}$~ & $1.60\times 10^{-2}$~ & $5.16\times 10^{+0}$~\\[0.5ex]\hline
\end{tabular}
\caption{Input parameters  for the linear response model at two rational surfaces in the pedestal of DIII-D discharge
\#158115. $m$ is the poloidal mode number, $n$ the toroidal mode number, $\hat{r}_s$ the rational surface radius normalized
to the plasma minor radius, 
$\epsilon\equiv (\epsilon_a/q_s)^{\,2}$, $\nu_\theta$ the  dimensionless poloidal flow  damping parameter, $\nu_\phi$ the dimensionless toroidal flow-damping parameter, 
$\nu_\mu$ the  dimensionless perpendicular viscosity parameter, $L$ the dimensionless locking parameter, and
$\hat{\omega}_0$ the normalized natural frequency. The latter quantity is calculated assuming that $\omega_0=-n\,(\omega_E+\omega_{\ast\,e})$.
 }\label{table2}
\end{table}

\begin{table}[p]
\begin{tabular}{cccccccc}\hline
$m$~ & ~$n$& ~$\nu_{\ast\,e}$~ & ~$\nu_{\ast\,i}$~ & $~\tau_{\rm SC}$~& ~$\delta_{\rm SC}$~ & $~\rho_s$~ & $b_f/b_v$ \\[0.5ex]\hline
-8 & $2$ & ~\,\,$9.26\times 10^{-2}$ & ~$2.97\times 10^{-2}$ &  ~$1.86\times 10^{-3}$~ & $4.14\times 10^{-3}$~ & ~$2.79\times 10^{-3}$~ & $1.45\times 10^{+1}$\\[0.5ex]
-11 &  $2$ & ~$3.80\times 10^{+0}$ & ~$1.22\times 10^{+0}$ &  ~$1.14\times 10^{-4}$~ & $5.75\times 10^{-3}$~ & ~$8.16\times 10^{-4}$~ &  $1.77\times 10^{+0}$\\[0.5ex]\hline
\end{tabular}
\caption{Important physical parameters at two rational surfaces in the pedestal of DIII-D discharge
\#158115. $m$ is the poloidal mode number, $n$ the toroidal mode number, $\nu_{\ast\,e}$ the dimensionless electron collisionality parameter,  $\nu_{\ast\,i}$ the dimensionless ion collisionality parameter, $\tau_{\rm SC}$ the semi-collisional reconnection timescale (s), $\delta_{\rm SC}$ the semi-collisional layer width (m), $\rho_s$ the ion sound radius (m), and $b_f/b_v$ the ratio of the normalized radial magnetic field to the
vacuum radial magnetic field at the rational surface. }\label{table3}
\end{table}

\begin{table}[p]
\begin{tabular}{ccccccccc}\hline
$m$~ & ~$n$~ & ~$\hat{r}_s$~& ~$\epsilon$~& ~$\nu_\theta$~  &~$\nu_\phi$~ & $\nu_\mu$ & $L$ &
 $\hat{\omega}_0$ \\[0.5ex]\hline
-8 & $2$ & ~$0.853$ & ~$1.77\times 10^{-2}$~ & $4.35\times 10^{+2}$~ & $0.0$ &  $2.16\times 10^{-3}$~ & $9.41\times 10^{-3}$~ & $6.97\times 10^{+1}$\\[0.5ex]
-11 & $2$ & ~$0.974$ & ~$9.34\times 10^{-3}$~ & $6.50\times 10^{+1}$~ & $0.0$ & $1.32\times 10^{-4}$~ & $1.60\times 10^{-2}$~ & $-9.72\times 10^{-1}$~\\[0.5ex]\hline
\end{tabular}
\caption{Input parameters  for the nonlinear response model at two rational surfaces in the pedestal of DIII-D discharge
\#158115. $m$ is the poloidal mode number, $n$ the toroidal mode number,  $\hat{r}_s$ the rational surface radius normalized
to the plasma minor radius, 
$\epsilon\equiv (\epsilon_a/q_s)^{\,2}$, $\nu_\theta$  the  dimensionless poloidal flow  damping parameter, $\nu_\phi$ the
dimensionless toroidal flow-damping parameter, 
$\nu_\mu$ the  dimensionless perpendicular viscosity parameter, $L$ the dimensionless locking parameter, and
$\hat{\omega}_0$ the normalized natural frequency. The latter quantity is calculated assuming that $\omega_0$ is given by (\ref{a14})--(\ref{a16}).
 }\label{table2a}
\end{table}

\section*{References}

\begin{thebibliography}{99}\baselineskip 5ex
\bibitem{wagner} Wagner F {\it et al.}\/ 1982 {\it Phys.\ Rev.\ Lett.}\/ {\bf 49} 1408 

\bibitem{zohm} Zohm H 1996  {\it Plasma Phys.\ Control.\ Fusion} {\bf 38} 105

\bibitem{den} Den Harden N {\it et al.}\/ 2016 {\it Nucl.\ Fusion} {\bf 56}, 026014  

\bibitem{loarte} Loarte A {\it et al.}\/ 2003 {\it J.\ Nucl.\ Materials} {\bf 313}--{\bf 316} 962

\bibitem{evans}  Evans T  E {\it et al.}\/ 2004 {\it Phys.\ Rev.\ Lett.}\/ {\bf 92} 235003

\bibitem{jet} Liang Y {\it et al.}\/ 2007  {\it Phys.\ Rev.\ Lett.}\/ {\bf 98} 265004 

\bibitem{asdex} Suttrop W {\it et al.}\/ 2011 {\it Phys.\ Rev.\ Lett.}\/{\bf 106} 225004 

\bibitem{kstar}  Jeon Y M {\it et al.}\/ 2012 {\it Phys.\ Rev.\ Lett.}\/  {\bf 109} 035004 

\bibitem{east} Sun T {\it et al.}\/ 2016 {\it  Phys.\ Rev.\ Lett.}\/  {\bf 117} 115001

\bibitem{conner} Connor  J W {\it et al.}\/ 1998 {\it Phys.\ Plasmas}\/ {\bf 5} 2687 

\bibitem{fenstermacher} Fenstermacher M E {\it et al.}\/ 2008  {\it Phys.\ Plasmas} {\bf 15} 056122

\bibitem{berc} B\'{e}coulet M {\it et al.}\/  2012 {\it Nucl.\ Fusion} {\bf 52} 054003


\bibitem{d158115} Nazikian R {\em et al.} 2015 {\it Phys.\ Rev.\ Lett.}\/ {\bf 114} 105002

\bibitem{orain} Orain F {\em et al.} 2019 {\it Phys.\ Plasmas}\/ {\bf 26} 042503

\bibitem{hu} Hu Q M {\it et al.}\/ 2019 
{\em Density dependaence of edge-localized-mode suppression and pump-out by resonant magnetic perturbations in the DIII-D tokamak}\/
Submitted to Phys.\ Rev.\ Lett.\

\bibitem{tm1} Yu Q, G\"unter G,  and Scott B D 2003 {\it Phys.\ Plasmas} {\bf 10} 797

\bibitem{tm2} Yu Q 2010 {\it Nucl.\ Fusion} {\bf 50} 025014

\bibitem{tm3} Yu Q, and G\"unter S 2011 {\it Nucl.\ Fusion} {\bf 51} 073030

\bibitem{rfw} Fitzpatrick R, and Waelbroeck F L 2005 {\it Phys.\ Plasmas} {\bf 12} 022307

\bibitem{colef} Cole A J, and Fitzpatrick R 2006 {\it  Phys.\ Plasmas} {\bf 13} 032503

\bibitem{rfx} Fitzpatrick R 2014 {\it Phys.\ Plasmas} {\bf 21} 092513 

\bibitem{rfy} Fitzpatrick R 2018  {\it Phys.\ Plasmas} {\bf 25}  082513 

\bibitem{rff} Fitzpatrick R 2018  {\it Phys.\ Plasmas} {\bf 25} 112505

\bibitem{rf1} Fitzpatrick R 1993 {\it Nucl.\ Fusion} {\bf 33} 1049 

\bibitem{fkr} Furth H P,  Killeen J, and Rosenbluth M N 1963 {\it Phys.\ Fluids} {\bf 6} 459 

\bibitem{ruth} Rutherford P H 1973 {\it Phys.\ Fluids}  {\bf 16} 1903 

\bibitem{drake} Drake J F, and Lee Y C 1976 {\it Phys.\ Fluids} {\bf 20} 1341

\bibitem{wael} Waelbroeck F L 2003 {\it Phys.\ Plasmas} {\bf 10} 4040 

\bibitem {transp} Hawryluk R J 1980 {\em Physics of Plasma Close to Thermonuclear Conditions Volume~1 }\/ (Commission of the European Communities, Brussels, Belgium) 19 Internal Document DUR-FU-BRU-XII/476180 

\bibitem{rf2} Fitzpatrick R 1998 {\it Phys.\ Plasmas} {\bf 5} 3325

\bibitem{hirsh} Hirshman S P 1978 {\it Nucl.\ Fusion} {\bf 18} 917 

\bibitem{stix} Stix T H 1973 {\it Phys.\ Fluids} {\bf 16} 1260

\bibitem{shaing} Shaing K 2003 {\it Phys.\ Plasmas} {\bf 10} 1443 

\bibitem{cole} Cole A J,  Hegna C C, and Callen J D 2008 {\it Phys.\ Plasmas} {\bf 15} 056102 

\bibitem{chapman} Chapman B E {\it et al.}\/ 2004 {\it Phys.\ Plasmas} {\bf 11} 2156 

\bibitem{abram1}  Abramowitz M, and Stegun I A (eds.) 1965 {\em Handbook of Mathematical Functions with
Formulas, Graphs, and Mathematical Tables}\/ (Dover, New York NY) Chapter~9

\bibitem{grad2} Gradshteyn I S, and Ryzhik I M 1980 {\em Table of Integrals, Series, and Products, Corrected and Enlarged Edition} (Academic Press, New York NY) Equation~6.521.1 

\bibitem{hel} Fitzpatrick R 1995 {\it Phys.\ Plasmas} {\bf 2} 825

\bibitem{paz1} Paz-Soldan C 2019 {\em private communication}

\bibitem{nazz} Nazikian R {\it et al.}\/ 2018 {\it Nucl.\ Fusion} {\bf 58} 106010 
  

\bibitem{paz} Paz-Soldan C {\it et al.}\/ 2019 {\it Nucl.\ Fusion} {\bf 59} 056012

\bibitem{omega1} La Haye R J {\it et al.}\/ 2003 {\it Phys.\ Plasmas} {\bf 10} 3644

\bibitem{omega2} Buratti P {\it et al.}\/ 2016 {\it Nucl.\ Fusion} {\bf 56} 076004  
 

\bibitem{four} Hazeltine R D, Kotschenreuther M, Morrison P J 1985 {\it Phys.\ Fluids} {\bf 28} 2466 

\bibitem{semi} Fitzpatrick R 1989 {\it Phys.\ Fluids B} {\bf 1} 2381 

\bibitem{semi1} Connor J W,  Hastie R J,  Helander P 2017 {\it J.\ Plasma Phys.}\/ {\bf 83} 

\bibitem{liu} Liu Y Q {\it et al.}\/ 2012 {\it Phys.\ Plasmas} {\bf 19} 072509 

\bibitem{bai} Bai X, Liu Y Q, and Gao Z 2017  {\it Phys.\ Plasmas} {\bf 24} 102505

\bibitem{ion} Cowley S C, Kulsrud R M, and Hahm T S 1986 {\it Phys.\ Fluids} {\bf 29} 3230

\bibitem{tor} Fitzpatrick R 2017 {\it Phys.\ Plasmas} {\bf 24} 072506 

\bibitem{sat1} Thyagaraja A 1981 {\it Phys.\ Fluids} {\bf 24} 1716 

\bibitem{sat2} Escande D F,  and Ottaviani M 2004 {\it Phys.\ Lett.\ A} {\bf 323} 278

\bibitem{sat3} Hastie R J,  Militello F, and Porcelli F 2005 {\it Phys.\ Rev.\ Lett.}\/ {\bf 95} 065001

\bibitem{ionpz} Fitzpatrick R 2012 {\it Plasma Phys.\ Control.\ Fusion} {\bf 54} 094002

\bibitem{orbit} Hinton F L, and  Kim, Y B 1995 {\it Phys.\ Plasmas} {\bf 2} 159

\bibitem{pazs} Paz-Solden C {\it et al.} 2015 {\it Phys.\ Rev.\ Lett.} {\bf 114} 105001

\bibitem{sauter} Sauter O, Angioni C, and Lin-Liu Y R 1999 {\it Phys.\ Plasmas} {\bf 6} 2834

\bibitem{grob} Kim Y B, Diamond P H, and Groebner R J 1991 {\it Phys.\ Fluids B} {\bf 3} 2050

\bibitem{hirsh1} Hirshman S P 1978 {\it Phys.\ Fluids} {\bf 21} 1295

\bibitem{zarn} Zarnstorff M C {\it et al.}\/ 1990 {\it Phys.\ Fluids B} {\bf 2} 1852

\bibitem{sigmar} Hirshman S P, and Sigmar D J 1981 {\it Nucl.\ Fusion} {\bf 21} 1079

\bibitem{grob1} Kim Y B,  Diamond P H, and Groebner R J 1992 {\it Phys.\ Fluids B} {\bf 4} 2996

\bibitem{callen} Callen J D  2010 {\em Viscous forces due to collisional parallel stresses for extended MHD codes}\/ (Report
UW-CPTC 09-CR) Available from {\tt http://www.cae.wisc.edu/\textasciitilde callen}

\bibitem{plasma} Fitzpatrick R 2014 {\em Plasma Physics: An Introduction} (Taylor \& Francis, Abingdon UK)

\end{thebibliography}
\end{document}